\documentclass[aip,jcp,reprint]{revtex4-1}

\usepackage{color,soul}
\usepackage{xspace}
\usepackage{xcolor}
\usepackage{tablefootnote}
\usepackage[normalem]{ulem}

\usepackage{booktabs}

\usepackage{physics}
\usepackage[version=3]{mhchem}
\usepackage{graphicx}
\usepackage{dcolumn}
\usepackage{bm}

\usepackage[utf8]{inputenc}
\usepackage[T1]{fontenc}
\usepackage{mathptmx}
\usepackage{etoolbox}

\makeatletter
\def\@email#1#2{%
 \endgroup
 \patchcmd{\titleblock@produce}
  {\frontmatter@RRAPformat}
  {\frontmatter@RRAPformat{\produce@RRAP{*#1\href{mailto:#2}{#2}}}\frontmatter@RRAPformat}
  {}{}
}%
\makeatother

\begin{document}

\preprint{AIP/123-QED}

\title{Effects of Interaction Range on Fluid Multicriticality: A Computational Study of an Interconverting Lattice Model}

\author{Thomas J. Longo}
\affiliation{Institute for Physical Science and Technology, University of Maryland, College Park, MD 20742,USA}
\email{tlongo1@umd.edu}

\author{Sergey V. Buldyrev}%
\affiliation{ Department of Physics, Yeshiva University, New York, NY 10033, USA}
\affiliation{Department of Physics, Boston University, MA 02215, USA}

\author{Fr\'ed\'eric Caupin}
\affiliation{Institut Lumi\`ere Mati\`ere, Universit\'e Lyon 1, CNRS, Institut Universitaire de France, F-69622 Villeurbanne, France}

\author{Mikhail A. Anisimov}
\affiliation{Institute for Physical Science and Technology, University of Maryland, College Park, MD 20742,USA}
\affiliation{Department of Chemical and Biomolecular Engineering, University of Maryland, College Park, MD 20742,USA}

\date{\today}

\begin{abstract}
The range of intermolecular interactions plays a central role in determining the nature of phase behavior and critical phenomena. It is well established through studies of the Ising model that as interaction range increases, Monte Carlo simulations progressively approach meanfield predictions as the effects of critical fluctuations are suppressed. In this work, we investigate how varying interaction range influences fluid multicriticality using an interconverting lattice model that exhibits both Ising-like liquid-gas criticality and symmetric fluid tricriticality (similar to that in the superfluid $^4$He-$^3$He mixture). This minimal polyamorphic-fluid model serves as a representative system for exploring the evolution of competing critical points within a generic framework. We analyze the model using both meanfield theory and three-dimensional Monte Carlo simulations while systematically varying the number of interacting neighbors, $Z_n$, from 6 to 388 on a simple cubic lattice. We find that the system with nearest-neighbor interactions ($Z_n=6$) reveals only two types of multicritical behavior, while for larger interaction ranges, four distinct archetypes emerge. We demonstrate the convergence of the simulation results to those of the meanfield theory as the number of interacting neighbors tends to infinity, and we discuss the results within the framework of crossover critical phenomena.
\end{abstract}


\maketitle 

\section{Introduction}~\label{Sec_Introduction}

Phase transitions are derived from the interactions between atoms or molecules. The correct treatment of these interactions represents a challenge to statistical physics. A powerful tool to address this challenge is the meanfield theory (MFT). By replacing the detailed interactions with an effective mean influence of the whole system on each of its particles, MFT captures the main features of phase transitions. However, its shortcomings are well known, in particular concerning critical behavior: MFT predicts that the critical exponents are the same in all dimensions, but these values differ from exact calculations and experiments~\cite{Nishimor_Critical_2011}. The MFT results become exact only above the upper critical dimension (\textit{e.g.} $d_\mathrm{c}=4$ for the Ising model), or for an infinite range of interactions~\cite{Stanley_Introduction_1971}.

One of the most detailed studies of the crossover from Ising criticality to meanfield (MF) criticality is found in works by Luijten and Binder~\cite{Luijten_Medium_1996,Luijten_Nature_1998,Luijten_Critical_1999,Binder_Crossover_2001}. Using large-scale Monte Carlo (MC) simulations of three-dimensional ($3d$) Ising models with tunable interaction ranges, they systematically investigated how critical behavior evolves from the asymptotic $3d$ Ising universality class to meanfield behavior as the effective interaction range increases. By studying equivalent-neighbor and finite-range Ising models over a broad set of interaction radii, they directly evaluated the effective critical exponents. They showed that while the true asymptotic exponents approach either the $3d$ Ising values or the MF values in the corresponding limits, the effective exponents observed at intermediate distances from the critical point vary continuously. Moreover, they show that the critical temperature decreases continuously upon increasing the range of interactions. An excellent agreement between the simulation results of Luijten and Binder with the crossover theory of critical phenomena was demonstrated by Kim \textit{et al.}~\cite{kim_crossover_2003} Although these studies provide some of the most comprehensive numerical calculations of how Ising-like critical exponents and values of the critical temperature evolve, to the best of our knowledge, there is no equivalent treatment for other types of universality classes or for multicriticality (except for a qualitative discussion of a symmetric binary mixture with a hard-core Yukawa potential~\cite{Paschinger_binary_2005}) that may exist in Ising-like systems with competing order parameters. It is also well known that the upper critical dimension for tricritical systems is $d_c=3$, meaning that the critical exponents are MF and the effects of fluctuations are reduced only to logarithmic corrections to the MF behavior~\cite{Riedel_Tricritical_1972,Wegner_Logarithmic_1973,Fisher_RG_1975,Stephen_Logarithmic_1975}. However, the crossover from pure MF behavior, expected to be observed at an infinite range of interactions, to logarithmic corrections for systems with a finite-range of interactions has not been studied so far.

In this work, we address these  questions through the study of a simple binary lattice with interconversion of species, the blinking checkers (BC) model, which is thermodynamically equivalent to a single-component system. It was originally introduced~\cite{Caupin_Polyamorphism_2021} to describe fluid polyamorphism (FP). FP refers to the existence of multiple fluid-fluid phase transitions in a single-component substance.~\cite{Debenedetti_Water_1998,Stanley_Liquid_2013,Anisimov_Polyamorphism_2018,Tanaka_Liquid_2020} FP has been proposed as an explanation for the anomalies in the properties of supercooled water~\cite{Ponyatovskii_Second_1994,Ponyatovskii_Metastable_1998,Tanaka_Simple_2000,Stokely_Hydrogen_2010,Stanley_Liquid_2013,Sing_TwoState_2014,Palmer_Metastable_2014,Gallo_Water_2016,Biddle_Twostructure_2017,Caupin_Thermodynamics_2019,Cerdeirina_Water_2019,Duska_Water_2020,Tanaka_Liquid_2020,Caupin_Polyamorphism_2021,Shi_Anomalies_2021,Yu_Unified_2023}. The existence of a liquid-liquid transition in supercooled water has been demonstrated by simulations of water-like microscopic models~\cite{Poole_Water_1992,Sastry_SingularityFree_1996,Abascal_TIP4P_2005,Ciach_Simple_2008,Stokely_Hydrogen_2010,Abascal_Widom_2010,Holten_Water_2014,Palmer_Metastable_2014,Singh_Two_2016,Gallo_Water_2016,Gonzalez_Comprehensive_2016,Biddle_Twostructure_2017,Debenedetti_Second_2020} and supported by experiments~\cite{Mishima_Decompression_1998,Kim_Experimental_2020}. The BC model, depending on the various interaction parameters, may mimic the behavior of supercooled water and lead to FP~\cite{Caupin_Polyamorphism_2021}. FP in the BC model is caused by a coupling between a conserved order parameter (density) and a specific nonconserved order parameter (fraction of interconversion).

Here, we use a symmetric version of the BC lattice, referred to as the degenerate blinking checkers (DBC) model, that demonstrates both Ising-like liquid-gas critical points and symmetrical tricritical points~\cite{Anisimov_Degenerate_2025}. Similar multicritical phase behavior has been extensively investigated in Ising-like systems with coupled order parameters~\cite{Blume_He_1971,Tavares_Heisenberg_1995,Wilding_Tricritical_1995} and in symmetric binary mixtures~\cite{Furman_Global_1977,Wilding_Symmetrical_1998,Paschinger_TypeIV_2005,Kofinger_Symmetric_2006}. These studies have revealed a rich variety of critical and tricritical phenomena, including four archetypal phase-diagram topologies (Types I-IV), originally classified by Tavares \textit{et al.}~\cite{Tavares_Heisenberg_1995} However, despite extensive investigations of these multicritical systems, the influence of interaction range on their phase-diagram topologies and associated crossover from fluctuation-affected to MF criticality remained much less studied.

While the DBC model exhibits similar four archetypal multicriticality~\cite{Anisimov_Degenerate_2025}, it is unique among the previously discussed systems as it manifests multicritical behavior in a single-component polyamorphic fluid. The DBC model represents a minimal framework for investigating the interplay between the scalar order parameter (density) and the vector-like (interconversion) order parameter that gives rise to the four archetypal phase behaviors. Additionally, the simple lattice formulation enables a systematic investigation of the influence of interaction range, from nearest-neighbor interactions ($Z_n=6$) to extended interactions ($Z_n=388$), with comparisons to the corresponding MF description. Through this approach, we characterize the crossover from fluctuation-affected to MF criticality and identify a special characteristic of the nearest-neighbor system at $Z_n = 6$. In particular, we show that the Type II and Type IV archetypes are not realized for this system, but emerge with increasing interaction range ($Z_n > 6$). Particular attention is given to the comparatively less-explored Type IV behavior, which we investigate in more detail to elucidate the influence of interaction range on its multicritical topology.

The manuscript is organized as follows. Section~\ref{Sec_BC_Model} recalls the key definitions and parameters of the DBC model. Section~\ref{Sec_SimulationMethods} provides the details of the MC simulations we performed. The results are presented in Section~\ref{Sec_Results}, which include the various phase diagrams and the dependence of the transition-point temperature and density on the range of interactions. Section~\ref{Sec_Conclusion} gives a conclusion and an outlook.

\section{Blinking Checkers Model}~\label{Sec_BC_Model}
The BC model is a lattice model with interactions between neighbors and the possibility of interconversion, which has been investigated in Refs.~\cite{Caupin_Polyamorphism_2021,Longo_Interfacial_2023,Buldyrev_BCM_2024,Anisimov_Degenerate_2025}. Each lattice site $i$ can be in one of three states: empty ($s_i = 0$), occupied by a type 1 particle ($s_i = 1$), or occupied by a type 2 particle ($s_i = 2$). The number of empty sites ($n_0$) is fixed, while the populations of particles of types 1 and 2 ($n_1$ and $n_2$) vary through an interconversion reaction, such that $n_0 + n_1 + n_2 = n$. The density and molecular fractions are defined as $\rho = (n_1 + n_2)/n$, $x_1 = n_1/(n_1 + n_2)$, and $x_2 = n_2/(n_1 + n_2)=1-x_1$. The model contains five free parameters: the energy and entropy change of the interconversion reaction ($e$, $s$), and three interaction energies ($\omega_{11}$, $\omega_{22}$, $\omega_{12}$) between species. The Hamiltonian of the system is given by
\begin{equation}\label{Eqn_IntEn_perSite}
    H=e n_2 + \frac{1}{2} \sum_{i=1}^{n} \sum_{j\in\mathcal{L}_i}\epsilon(s_{i},s_{j})
\end{equation}
where $\mathcal{L}_i$ is the set of $Z_n$ neighbors of site $i$, and $\epsilon(s_i, s_j) = -2\omega_{ij}/Z_n$ in which $\omega_{ij} = \omega_{ji}$. Interactions involving empty sites are zero. More details about the general model can be found in Ref.~\onlinecite{Caupin_Polyamorphism_2021}.

In the degenerate blinking-checkers (DBC) model considered here, $e=0$ and $s=0$, and the self-interactions are symmetric ($\omega_{11} = \omega_{22}$), so that the equilibrium constant for interconversion satisfies: $\ln K = \mu_2^\circ -\mu_1^\circ = e - Ts = 0$, where $\mu_1^\circ$ and $\mu_2^\circ$ are the chemical potentials of pure species $1$ and $2$, respectively. In that case, the MF Helmholtz free-energy per lattice site is:
\begin{eqnarray}
&f(T,\rho,x_1 ) = - \rho^2 \left[ \omega_{11} -2 (\omega_{11} - \omega_{12}) x_1 x_2 \right]\nonumber\\
&+ T \left\{ \rho \left[ x_1 \ln x_1 + x_2 \ln x_2 \right] + \rho \ln \rho + (1-\rho) \ln (1-\rho) \right\}
\end{eqnarray}

In the following, we use the equations derived in the supplemental material of Ref.~\onlinecite{Caupin_Polyamorphism_2021}. Chemical-reaction equilibrium is enforced by setting $\partial f/\partial x|_T = 0$, where $f$ is the Helmholtz free energy per unit volume, which imposes a constraint on the equilibrium composition, $x_e=x_e(T,\rho)$, where $\rho$ is the number density. The condition $\ln{K}=0$ is equivalent to zero magnetic field in the Ising model, making the conjugate fraction of interconversion a non-conserved order parameter. The chemical potential is obtained as $\mu = \partial f/\partial \rho|_{T}$ and evaluated along the equilibrium concentration, $x=x_e$. Phase coexistence is determined by applying the conditions of equal pressure and chemical potential. Spinodal curves are obtained from $\partial \mu/\partial \rho|_T = 0$; their shape determines the location of the fluid-fluid critical point (where the first-order transition terminates), the tricritical point (where the first-order transition becomes second-order), and a $\lambda$-line (the line of the second order phase transitions) above which the concentration of the interconverting species is $x_1=x_2=1/2$. 

To compare with simulation data, the temperature is normalized as $\hat{T} = k_\text{B}T/\omega_{11}$ with Boltzmann's constant set to unity $k_\text{B}=1$. Moreover, the MF lambda line is defined by $\hat{T}_{\lambda}^\text{MF} = 2\rho_{\lambda}^\text{MF}(1-\omega_{12}/\omega_{11})$. This motivates the definition of a reduced interaction parameter, 
\begin{equation}\label{Eq_wBar}
    \bar{\omega} = 1-\frac{\omega_{12}}{\omega_{11}} \, ,
\end{equation}
so that the MF $\lambda$-line is given by $\hat{T}_{\lambda}^\text{MF} = 2\bar{\omega}\rho_{\lambda}^\text{MF}$. In the special case of an ideal solution ($\omega_{12} = \omega_{11} = \omega_{22}$, hence $\bar{\omega} = 0$), no liquid-liquid phase transition occurs. Only a liquid-gas transition is observed, with $\hat{T}_\text{c} = 1$ and $\rho_\text{c} = 0.5$.

\section{Simulation Methods}\label{Sec_SimulationMethods}

In order to construct the phase equilibrium diagrams of the DBC model, MC simulations were conducted on a simple cubic lattice in an elongated box with dimensions $\ell_x=\ell_y=128$ and $\ell_z=256$, giving a total of $n = \ell_x\ell_y\ell_z = 2\ell_x^3$ lattice sites. Periodic boundary conditions were applied in all directions.

\begin{figure}[tbp]
    \centering
    \includegraphics[width=0.99\linewidth]{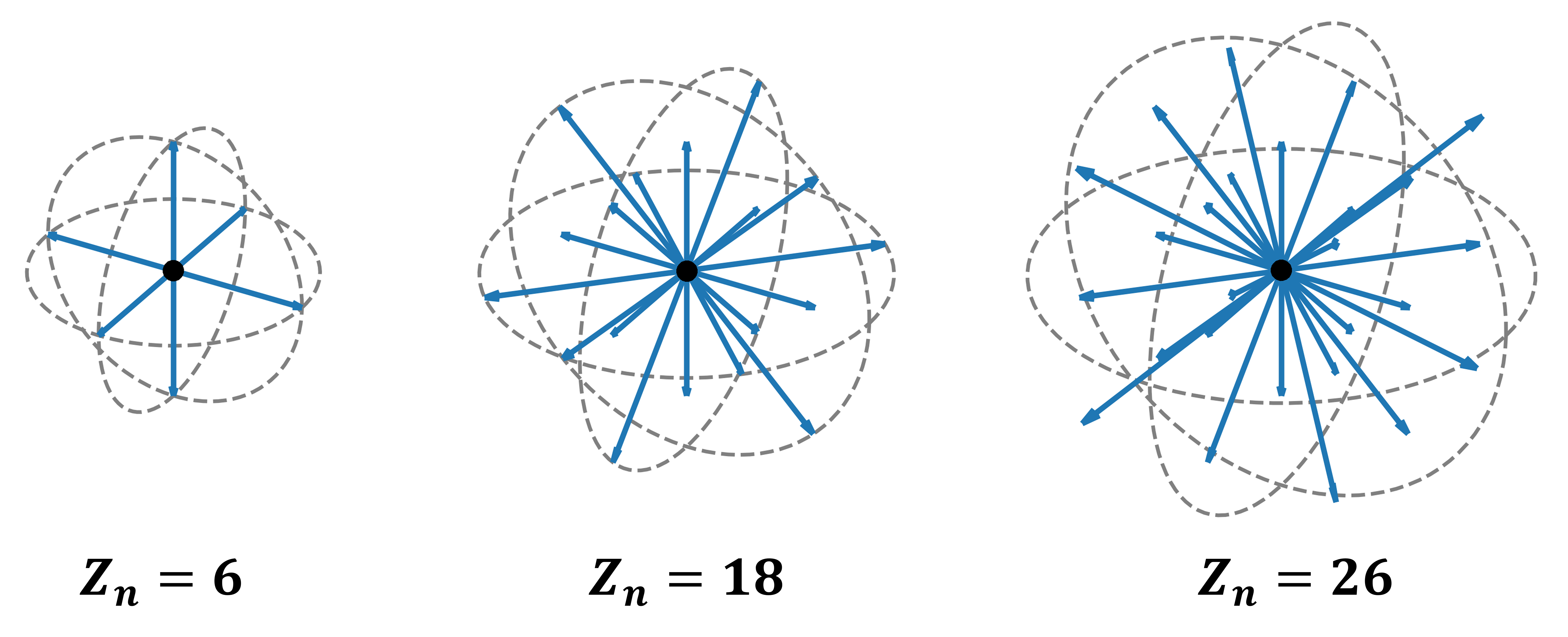}
    \caption{Depiction of the range of interactions between a given lattice site and its neighbors for the first three spherical shells of radii: $r_z = 1$ ($Z_n=6$), $r_z=\sqrt{2}$ ($Z_n=18$), and $r_z=\sqrt{3}$ ($Z_n=26$).}
    \label{Fig_Zn_Qual}
\end{figure}

The interaction range between two occupied sites is determined by their Euclidean distance under periodic boundary conditions. For a given cutoff radius $r_z$, the coordination number $Z_n$ is defined as the number of distinct lattice sites whose distance from a reference site lies within a spherical shell of radius $r$, such that $0< r\leq r_z$. The function $Z_n(r_z)$ is a non-decreasing step function. The simulations conducted in this work were performed for: $Z_n(1.1)=6$, $Z_n(1.45)=18$,  $Z_n(1.75)=26$,
$Z_n(2.45)=80$, $Z_n(3.1)=122$ , $Z_n(3.9)=250$, and $Z_n(4.5)=388$. 

\begin{figure}[b!]
    \centering
    \includegraphics[width=0.9\linewidth]{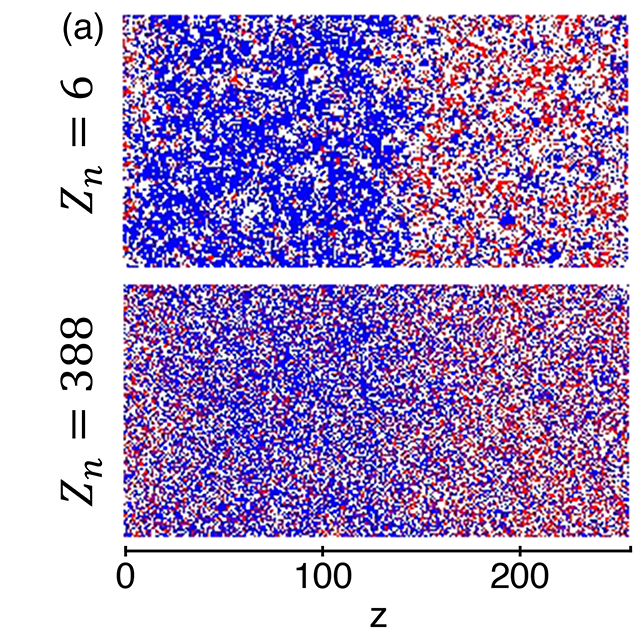}
    \includegraphics[width=0.9\linewidth]{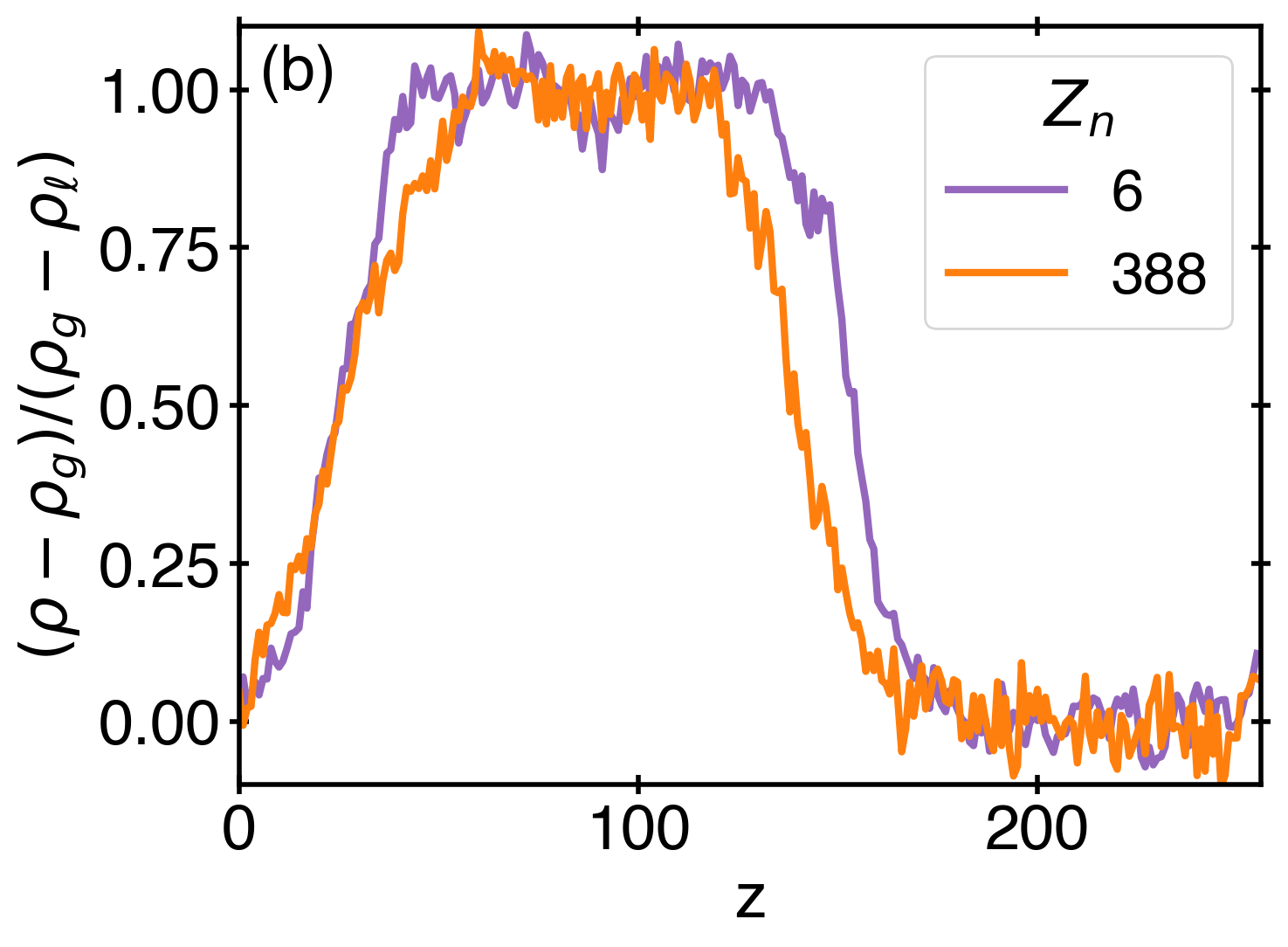}
    \caption{Phase coexistence obtained from MC simulations with $\bar{\omega}=1$ after $2^{33}$ elementary MC steps, corresponding to approximately 2048 update opportunities per lattice site, and averaged over eight independent realizations. Two systems are presented, depicting the smallest ($Z_n=6$ at $\hat{T}=0.8$) and largest ($Z_n=388$ at $\hat{T}=1.113$) coordination numbers investigated in this work. Both temperatures were selected such that the systems are approximately $0.1$ units below their respective critical points. Panel (a) shows a representative 2D slice of the simulation configurations in the $xz$-plane, with sites colored in white, blue, and red, if empty or occupied by particles of type 1 or 2, respectively. Panel (b) shows the corresponding density profiles along the $z$-direction for $Z_n=6$ (purple) and $Z_n=388$ (orange).}
\label{fig_profile_and_correlation_function}
\end{figure}

Each MC step consists of a Kawasaki (exchange) move between an empty and an occupied site~\cite{kawasaki_diffusion_1966} followed by a Glauber (interconversion) flip~\cite{glauber_timedependent_1963}. Note that while the energetic interactions in the system are computed over the range defined by $r_z$, Kawasaki exchanges are performed for any pair of occupied and empty lattice sites. The proposed moves are accepted according to the Metropolis criterion~\cite{metropolis_basic_1963}, in which a move is always accepted when $\Delta F < 0$ and is accepted with probability, $P = \exp(-\Delta F/T)$ when $\Delta F > 0$. The change in the free energy is calculated as $\Delta F = \Delta U - T\Delta S$ where the total potential energy is
\begin{equation}\label{eq:UMC}
    U=\frac{1}{2}\sum_i^n\sum_{j\in\mathcal{L}_i}\epsilon(s_i,s_j)+\tilde{e} n_2
\end{equation}
For Kawasaki exchanges, $\Delta S = 0$, while for Glauber flips, $\Delta S = \pm \tilde{s}$, where $\tilde{s}$ is the internal entropy of the interconversion reaction. The positive sign corresponds to conversion from type 1 to type 2, while the negative sign corresponds to conversion from type 2 to type 1. In this work, we consider the degenerate case $\tilde{e} = \tilde{s} = 0$.

In each Glauber step, a random site is selected and allowed to flip according to the Metropolis criterion. In each Kawasaki step, a pair is selected randomly from the list of available empty sites and from the list of filled sites and allowed to exchange according to the Metropolis criterion. Following each successful flip or exchange, the lists of empty and occupied sites are updated. Consequently, a size-independent MC time is defined as $t = N_\text{MC}/(\ell_x\ell_y\ell_z)$ where $N_\text{MC}$ is the cumulative number of effective elementary MC steps. This normalization ensures that each lattice site has the same average opportunity to undergo one flip and one exchange per unit time. All simulations were conducted for at least $2^{33}$ elementary MC steps, corresponding, on average, to 2048 update attempts per lattice site. This number of MC steps was selected because it is greater than the equilibration time of the system.

The result of these interconversion dynamics is the phenomenon of phase amplification, in which one liquid phase grows at the expense of the other to avoid the formation of an energetically unfavorable interface~\cite{Shum_Phase_2021}. Although the present work focuses exclusively on equilibrium phase behavior, the non-equilibrium dynamics associated with phase amplification in these systems is an interesting topic for future study.

To promote the formation of two coexisting phases separated by two interfaces (arising from periodic boundary conditions) perpendicular to the elongated $z$-direction, simulations were initialized with all sites in the region $z < \rho\ell_z$ occupied by type 1 particles, while the remaining sites were left empty. For each state point, eight independent configurations were recorded at time intervals of $2^{33}$ MC steps. Only the final equilibrium configuration from each realization was analyzed, and all reported properties were obtained by averaging these eight realizations. If the selected state point ($T$,$\rho$) lies below the $\lambda$-line, the system remains in a state in which only the phase (liquid or gas) with $x_1 >x_2$ survives due to phase amplification, and the second phase will never form. Above the $\lambda$-line the concentrations of the two types of particles rapidly equilibrate to $x_1=x_2=0.5$. Note that in archetype I (small $\bar\omega$), the gas phase is always disordered $x_1=x_2=0.5$, whereas in archetype IV, the gas phase may become ordered ($x_1>x_2$). For each state point, to determine the equilibrium density and concentration along phase boundaries, we compute density and composition profiles along the $z$ axis, identify their extrema, and calculate their average values within  windows of $\Delta z=64$ centered on the positions of minimum and maximum density. An example of profiles is presented in Fig.~\ref{fig_profile_and_correlation_function} along with simulation snapshots for the smallest ($Z_n=6$) and largest ($Z_n=388$) coordination numbers considered in this work. As expected, the larger coordination number, the broader the liquid-vapor interface.

To determine the location of the $\lambda$-line, we computed histograms of the order parameter $\varphi=|x_1-1/2|$ for selected state points above the coexistence line in the $T$-$\rho$ plane. Simulations were run for 10 times longer than the equilibration time. Histogram bins spanned the range from $\varphi=0$ to the maximum $\varphi$ observed during the simulation and were divided into 10 equal bins. Following the criterion established by Binder~\cite{Binder_finite_1981} and Bruce~\cite{Bruce_probability_1981}, the second-order transition temperature in a finite-sized system is identified by a bimodal order-parameter distribution for which the probability density at the minimum ($\varphi=0$) is one-half of that at the maxima. For each fixed $\rho$, simulations were performed with a small temperature increment, and the second-order transition ($\lambda$-line) temperature was determined by interpolating between temperatures at which the histogram minimum at $\varphi=0$ crossed one-half of the histogram maximum.

\section{Results and Discussion}~\label{Sec_Results}

The results of our study are separated into two parts. In Section~\ref{Sec_Phenomenology}, we review the phenomenology of the four archetypal systems considered in this work and in the literature, while in Section~\ref{Sec_CoordinationNumber}, we present the critical, tricritical, and $\lambda$-end-point temperatures and densities quantitatively as a function of the coordination number, $Z_n$.

\begin{figure*}[th!]
	\centering
	\includegraphics[width=0.49\linewidth]{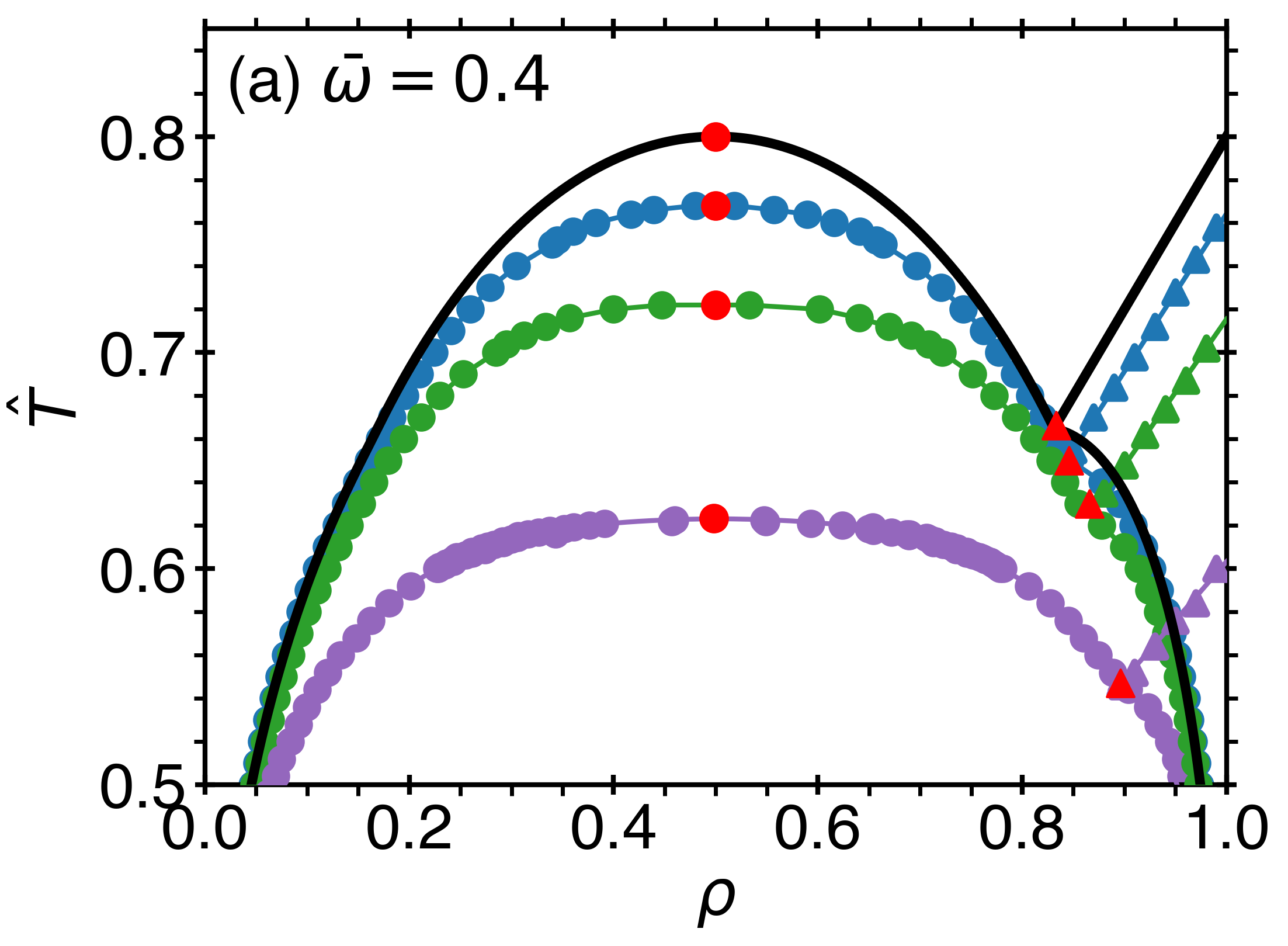}
	\includegraphics[width=0.49\linewidth]{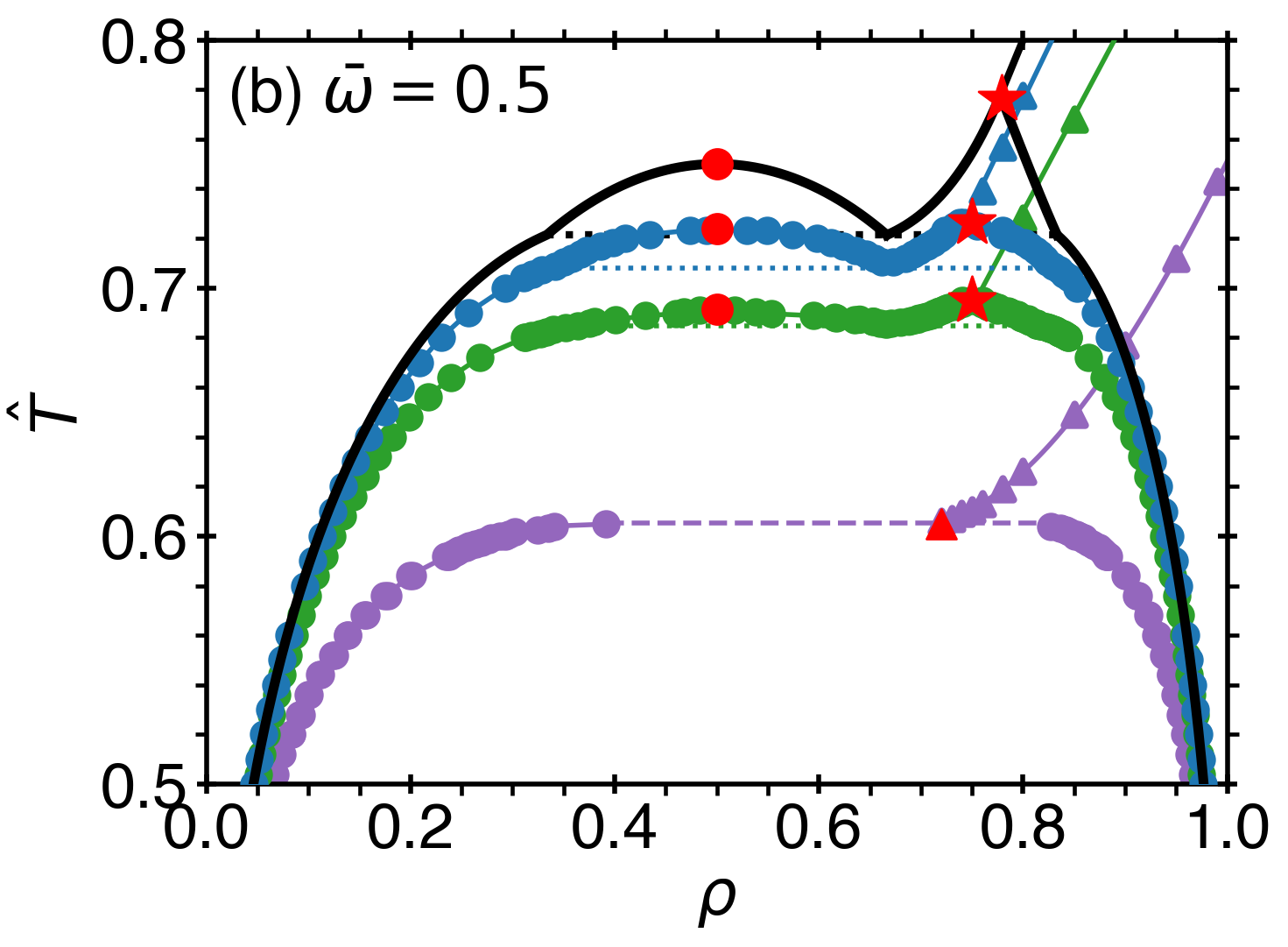}
	\includegraphics[width=0.49\linewidth]{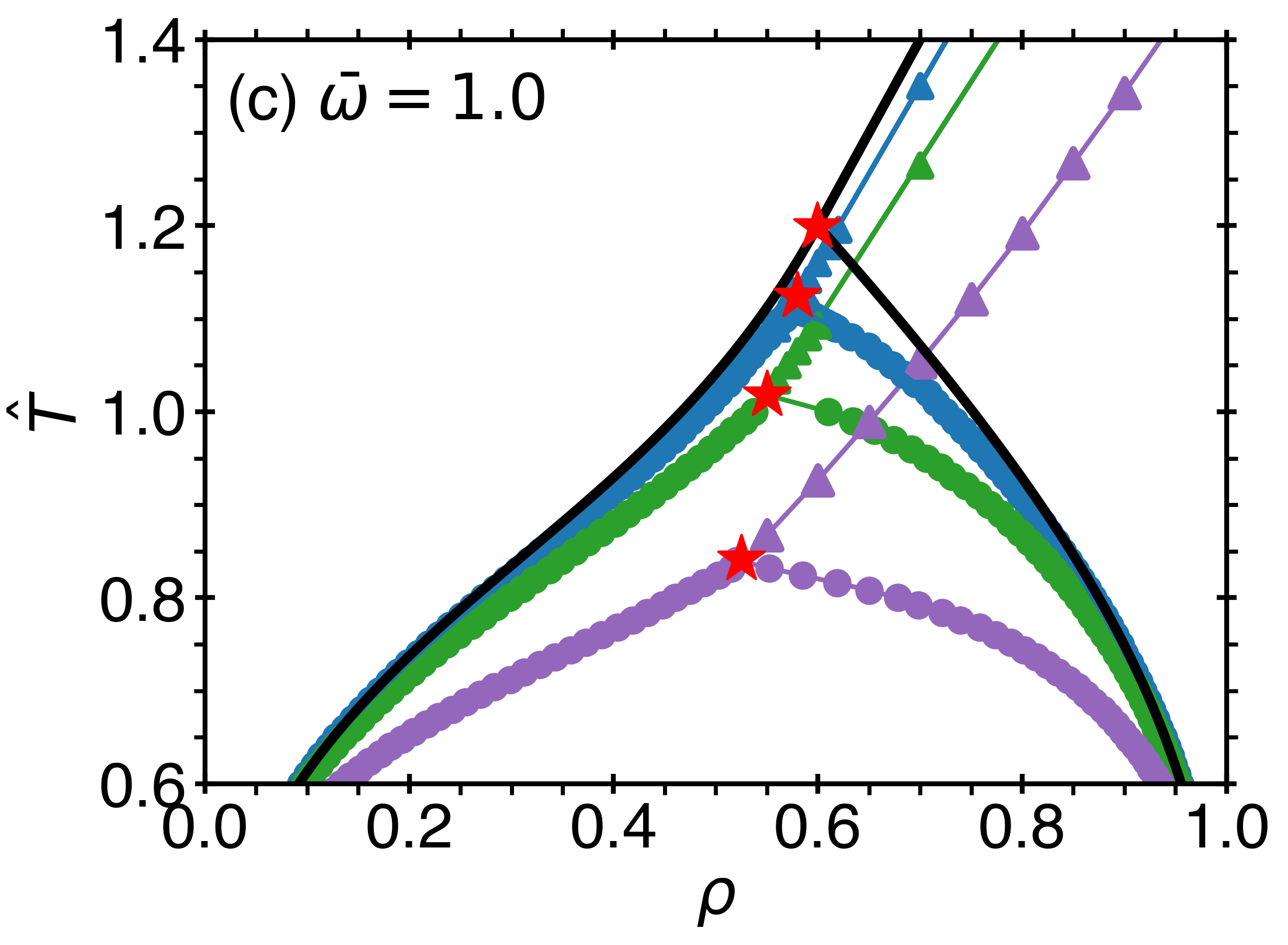}
	\includegraphics[width=0.49\linewidth]{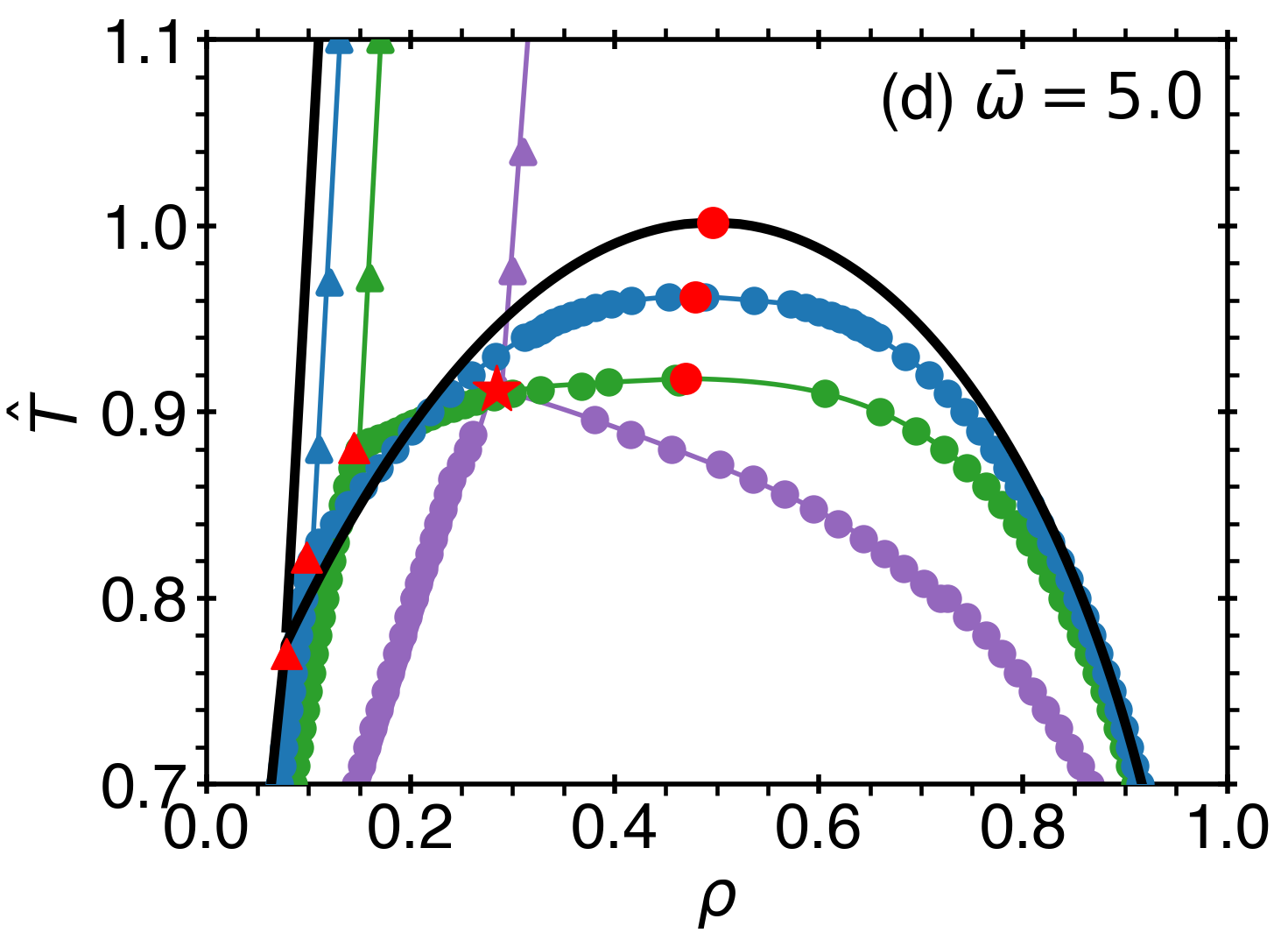}
	\caption{Temperature-density phase diagrams for four representative values of $\bar{\omega}$ (a-d) shown for three coordination numbers: $Z_n=6$ (purple), $Z_n=26$ (green), and $Z_n=80$ (blue), together with the MF theory predictions (black). Note that, for clarity, only these three coordination numbers are shown; additional $Z_n$ values considered in this work exhibit the same overall trends, which are considered in more detail in the next Section (\ref{Sec_CoordinationNumber}). In all panels, circles denote liquid–gas coexistence data, while triangles indicate liquid–liquid or vapor-vapor second-order phase transition ($\lambda$-line) data. The thin curves are provided as guides to the eye. The dashed line in (b) represents the discontinuity in the first-order transition (more details in Appendix~\ref{Sec_Appendix_SpecialCase}). Red symbols identify key thermodynamic points: critical points (red circles), $\lambda$-end points (red triangles), and tricritical points (red stars). The four selected values of $\bar{\omega}$ generally correspond to archetypes: $\bar{\omega}=0.4$ (I), $\bar{\omega}=0.5$ (II), $\bar{\omega}=1.0$ (III), and $\bar{\omega}=5.0$ (IV); however, the systems with $Z_n=6$ do not exhibit Type II or Type IV behavior. Accordingly, the $Z_n=6$ data shown in panels (b) and (d) correspond to Type I and III behavior, respectively.}
	\label{Fig_cxc}
\end{figure*}

\begin{table*}[th!]
	\caption{\label{tab_MF_critVals} MF values for the temperature and density at the critical, tricritical, or lambda end point (if any) for each $\bar{\omega}$ investigated, illustrating the unique behavior of the four archetypes~\cite{Anisimov_Degenerate_2025}.}
	\begin{ruledtabular}
		\begin{tabular}{cccccccc}
			$\bar{\omega}$ & Archetype & $\hat{T}_\text{CP}^\text{MF}$ & $\rho_\text{CP}^\text{MF}$ & $\hat{T}_\text{TCP}^\text{MF}$ & $\rho_\text{TCP}^\text{MF}$ & $\hat{T}_{\lambda \text{EP}}^\text{MF}$ & $\rho_{\lambda \text{EP}}^\text{MF}$ \\ \hline
			0.4 & I                     & 0.800 & 0.500                         & - & -           & 0.665 & 0.831             \\
			0.5 &II                    & 0.750 & 0.500                         & 0.777 & 0.777   & - & -             \\
			1.0 & III                    & -     & -                             & 1.200 & 0.600   & - & -             \\
			5.0 & IV                     & 1.002 & 0.496                         & - & -           & 0.776 & 0.078  
		\end{tabular}
	\end{ruledtabular}
\end{table*}

\subsection{Phenomenology of Four Archetypes}\label{Sec_Phenomenology}

In this subsection, we review the temperature-density phase diagrams of the four systems considered in this work and compare the DBC model to other multicritical models present in the literature.

\subsubsection{Phase Diagrams}\label{Sec_PhaseDiagrams}

Figure~\ref{Fig_cxc} presents temperature-density phase diagrams, obtained from MC simulations at four values of $\bar{\omega}$ (given by Eq.~\ref{Eq_wBar}) chosen to illustrate each of the four archetypes: 0.4 (I), 0.5 (II), 1 (III), and 5 (IV). For clarity, only three representative coordination numbers ($Z_n = 6$, $Z_n = 26$, and $Z_n = 80$) out of the seven values studied in this work are shown, together with the corresponding MF predictions (black curves). As shown in Fig.~\ref{Fig_cxc}, Type I critical behavior features both a conventional liquid-gas critical point (LGCP) and a lambda line ($\lambda$-line) that intersects the liquid branch of the coexistence curve and terminates at a lambda-end point ($\lambda$EP). Type II critical behavior is characterized by the presence of two critical points: an LGCP and a symmetrical tricritical point (TCP). Type III critical behavior exhibits only a TCP. In Types II and III, the $\lambda$EP coincides with the TCP. Type IV critical behavior, similar to Type I, features a conventional LGCP, but the $\lambda$-line intersects the vapor branch of the coexistence curve, indicating that a lambda transition can occur in the gaseous phase.

The CP and TCP were determined from the intersections of extrapolations of coexistence data for the gaseous and liquid branches to the respective transition points. For CP, both branches were extrapolated with curved fits, whereas for the TCP the gaseous branch was assumed to be linear and the liquid branch curved. The $\lambda$EP was obtained from the extrapolated intersection of the $\lambda$-line with the liquid branch for Type I systems and the gaseous branch for Type IV systems. Based on this extrapolation procedure, uncertainties of approximately $\hat{T}\pm 0.0025$ and $\rho\pm 0.01$ were assigned to the estimated transition-point temperature and density locations, respectively.

The four archetypes were systematically examined across a range of coordination numbers. In all cases, increasing the coordination number (to larger $Z_n$) causes the locations of the key transition points, coexistence curves, and $\lambda$-lines to converge toward the MF predictions in the limit $Z_n \to \infty$. This general trend, caused by the progressive suppression of fluctuations as the interaction range increases, is consistent with the crossover theory developed for the three-dimensional Ising model~\cite{kim_crossover_2003} and has been observed in the equivalent-neighbor lattice model (see Ref.~\onlinecite{Luijten_Critical_1999} and references therein) and the $2d$ spin-fluid model~\cite{Casiulis_Ferromagnetisminduced_2019}. In particular, we note that for the TCP, deviations from the MF location are also observed as a function of $Z_n$, despite being simulated at the marginal dimension ($d_m=3$)~\cite{Riedel_Tricritical_1972}. This may be attributed to the presence of logarithmic corrections to the phase behavior~\cite{Lawrie_Tricritical_1984,Riedel_Tricritical_1972,Wegner_Logarithmic_1973,Fisher_RG_1975,Stephen_Logarithmic_1975}. A more detailed discussion of the dependence of these three key transition points on $Z_n$ is presented in Section~\ref{Sec_CoordinationNumber}. 

\subsubsection{Unique Behavior of the Nearest-Neighbor Model}\label{Sec_NN_DBC_Behavior}

\begin{figure}[t]
    \centering
    \includegraphics[width=\linewidth]{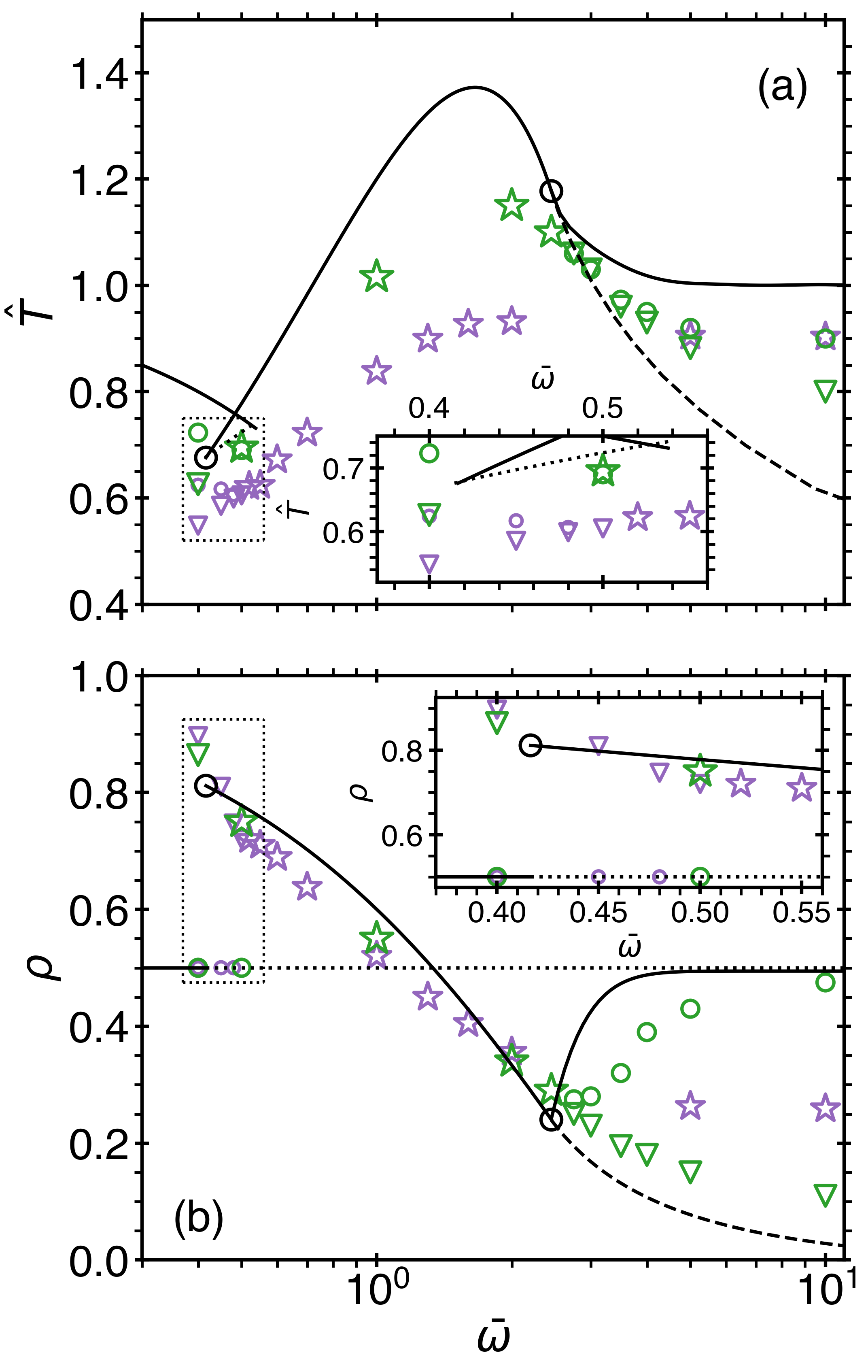}
    \caption{Effect of interaction range on the multicritical behavior of the DBC model, highlighting the loss of the special nearest-neighbor feature upon increasing the coordination number, $Z_n$, from $Z_n=6$ (purple symbols from Ref.~\onlinecite{Anisimov_Degenerate_2025}) to $Z_n = 26$ (green symbols; this work), shown as a function of the nonideality interaction parameter, $\bar{\omega}$, in terms of (a) temperature and (b) density. The symbols represent MC simulation results for liquid-gas critical points (circles), tricritical points (stars), $\lambda$-end points (triangles). Black curves and circles depict the MFT predictions reported by Furman \textit{et al.}~\cite{Furman_Global_1977} for a symmetric three-component lattice model without interconversion. Solid curves correspond to the CP and TCP loci, and the dashed curves show the $\lambda$EP locus. The circles mark the predicted locations of the Type I-II and Type II-III transitions at $\bar{\omega}=0.4165$ and $\bar{\omega}=2.4503$, respectively~\cite{Furman_Global_1977}. The dotted curve in (a) denotes the triple line, while in (b), it serves as a guide to the eye at $\rho=0.5$. Insets in (a) and (b) highlight the absence of Type II behavior in the nearest-neighbor system ($Z_n=6$) near the onset of Type III behavior.}
    \label{fig_Zn6_symmetry}
\end{figure}

Although, for each archetype, most coordination numbers yield phase diagrams that are qualitatively similar to the MF predictions, differing primarily through shifted critical parameters and the well-known modification of the critical exponents~\cite{Nishimor_Critical_2011,kim_crossover_2003,Cardy_Scaling_1996,Goldenfeld_Lectures_2018}, the nearest-neighbor system with $Z_n = 6$ exhibits the largest deviations. These departures arise from the geometry of the interacting neighbors and have been referred to as ``lattice effects''~\cite{Luijten_Nature_1998}. In our previous study~\cite{Anisimov_Degenerate_2025}, we investigated the multicritical behavior of the DBC model with nearest-neighbor interactions ($Z_n=6$) and presented preliminary results for extended interactions ($Z_n=26$), which suggested a qualitative change in the phase behavior. Here, we build on these preliminary findings by comparing the multicritical behavior of the $Z_n=6$ and $Z_n=26$ systems over a broader range of $\bar{\omega}$. Figure~\ref{fig_Zn6_symmetry} compares multicritical behavior of the two systems along with the corresponding MF predictions as functions of the nonideality parameter, $\bar{\omega}$, in (a) temperature and (b) density space.

Unlike the MF and $Z_n=26$ systems, the $Z_n=6$ system exhibits a direct transition from Type I to Type III behavior, without an intervening Type II regime. Consequently, the system with $\bar{\omega}=0.5$ (see Fig.~\ref{Fig_cxc}b) exhibits a unique $\lambda$EP at approximately $\rho=0.72$, where the $\lambda$-line meets a discontinuity in the first-order phase transition. In Appendix~\ref{Sec_Appendix_SpecialCase}, the nearest-neighbor DBC model at $\bar{\omega}=0.5$ is mapped to a symmetric three-state Potts model. This mapping establishes the existence of the discontinuity in the first-order phase transition and suggests that the absence of Type II behavior may be related to the interplay between the underlying symmetry and the restriction to nearest-neighbor interactions. We note that the point at which the $\lambda$-line meets this discontinuity warrants further investigation.

A similar distinction is observed for the Type IV archetype, where the tricritical line separates into the CP and $\lambda$EP loci in the MF and $Z_n=26$ systems, but no such separation occurs in the system with $Z_n=6$. In the MF description, this separation occurs at $\bar{\omega}=2.4503$,~\cite{Furman_Global_1977} whereas for $Z_n=26$, it occurs near $\bar{\omega}\approx 2.5$, slightly above the MF prediction. In both systems, the critical density reaches its maximum deviation from the symmetric value of $\rho_\text{c}=0.5$ at this boundary. For $Z_n=6$, however, the critical density approaches its maximum deviation only as $\bar{\omega}\to\infty$. In the system with $\bar{\omega}=5.0$ (Fig.~\ref{Fig_cxc}d), all cases with $Z_n > 6$ exhibit distinct CP and $\lambda$EP points, whereas the $Z_n=6$ phase diagram remains consistent with a single TCP, like that of the system with $\bar{\omega}=1.0$ (see Fig.~\ref{Fig_cxc}c). The apparent absence of Type IV behavior at $Z_n=6$ may arise from the strongly negative interaction parameter, $\omega_{12}$, required to access this regime, which may prevent the stabilization of the ordered gaseous phase associated with the intersection of the $\lambda$-line and the vapor branch of the coexistence curve. However, within the uncertainty of the simulation data, we cannot conclusively distinguish among a genuine TCP, closely spaced CP and $\lambda$EP points, and coincident CP and $\lambda$EP points, producing a confluence of critical behavior. Further investigation, including an analysis of second-order thermodynamic properties such as susceptibility or correlation length, is therefore required to establish whether the Type IV behavior is genuinely absent at $Z_n=6$ and to clarify the physical origin of this behavior.

\subsubsection{Comparison with other multicritical models}~\label{Sec_LiteratureComparison}

Similar symmetric multicritical phase behavior has been observed in two general classes of systems: (i) Systems with symmetric (``degenerate'') interconversion of species or states, such as Ising-like lattice models with ``non-magnetic'' impurities and spin-fluid models, and (ii) non-interconverting systems with a special symmetry, including a symmetric compressible binary (SCB) lattice and quasi-binary ternary-lattice mixtures.

The first class includes the Blume-Capel and Blume-Emery-Griffiths (BEG) models, which describe Ising-like systems with a nonmagnetic impurity~\cite{Blume_BC_1966,Capel_BC_1966,Blume_He_1971,Moueddene_Blume_2024}. The BEG model was originally developed to describe phase separation and superfluid ordering in $^3$He-$^4$He mixtures and was later generalized~\cite{Berker_VectorBEG_1979,Freire_VectorBEG_2005}. Under special symmetry conditions, it is also equivalent to the three-state Potts model~\cite{Berker_BEG_Potts_1976,Antenucci_BEG_2014}, providing a connection to the nearest-neighbor DBC model at $\bar{\omega}=0.5$ (see Appendix~\ref{Sec_Appendix_SpecialCase}). Related continuum models include compressible spin fluids with spin-dependent interactions~\cite{Hoye_SpinSystem_1976}, for which Wilding and Nielaba demonstrated tricritical behavior in two dimensions~\cite{Wilding_Tricritical_1995}. Further investigations of continuum Heisenberg fluids by Lomba \textit{et al.}~\cite{Lomba_Heisenberg_1994} using MC simulations and Tavares \textit{et al.}~\cite{Tavares_Heisenberg_1995} using density-functional theory revealed the rich phase behavior arising from the coupling between magnetic ordering and liquid-vapor transitions. In particular, Tavares \textit{et al.}~\cite{Tavares_Heisenberg_1995} defined the four archetypes (I-IV) of critical phase behavior adopted in our study.

The second class comprises SCB mixtures and incompressible quasi-binary ternary mixtures without the possibility to exhibit interconversion. Furman \textit{et al.} investigated the global phase behavior of a three-component quasi-binary incompressible lattice model in the meanfield approximation.~\cite{Furman_Global_1977} Wilding \textit{et al.} used MF theory and MC simulations to identify Type I, II, and III behavior in SCB square-well fluids~\cite{Wilding_Symmetrical_1998}. Sch\"oll-Paschinger and Kahl later demonstrated Type IV behavior in a SCB fluid with a hard-core Yukawa (HCY) potential~\cite{Paschinger_TypeIV_2005}. Subsequently, K\"ofinger \textit{et al.}~\cite{Kofinger_Symmetric_2006} investigated the Type II regime of a SCB HCY fluid using the mean-spherical approximation (MSA) and grand-canonical MC simulations. They also observe two distinct phase-diagram subtypes upon symmetry breaking of the chemical potentials of the two species~\cite{Kofinger_Symmetric_2006}.

The DBC model combines characteristics of the two previously discussed classes, representing Ising-like degenerate interconversion of species in a SCB lattice. Consequently, such a binary system behaves like a single-component substance in which the total density is conserved but the fraction of interconversion is not. In the DBC model, the role of a ``non-magnetic impurity'' is played by empty spots in the lattice, while the interconversion order parameter behaves like the magnetization in zero magnetic field. Consequently, due to the non-conserved order parameter the phase coexistence of two alternative ``magnetic'' phases and the formation of interfaces are suppressed by the phenomenon of phase amplification~\cite{Shum_Phase_2021}. As a result, the line of transitions in $P$-$T$ space between two fluid phases with the same density, ``disordered'' (50/50 interconverted) and ``ordered'' (partially interconverted), becomes equivalent to the $\lambda$-line in liquid $^4$He. We note that to achieve symmetric multicriticality, the binary system needs to be ``fully degenerate,'' exhibiting both a symmetry in its component properties and a reduction of the number of field variables by either degenerate interconversion or fixing the component concentrations to be equal.

Another example of a SCB fluid with degenerate interconversion of species is a mixture of two interconverting enantiomers (chiral isomers). A model for such a system was investigated by molecular dynamics simulations~\cite{Latinwo_MolecModel_2016,Uralcan_Interconversion_2020,Petsev_Effect_2021,Wang_Chiral_2022}, and is referred to as the ``chiral model.'' In this model, the kinetic barrier of interconversion is controlled by a rigidity parameter, while unfavorable interaction between the isomers is controlled  by a ``bias parameter,'' analogously to the parameter $\bar{\omega}$ in the DBC model. Indeed, as shown in Ref.~\onlinecite{Latinwo_MolecModel_2016,Uralcan_Interconversion_2020,Petsev_Effect_2021,Wang_Chiral_2022}, at low values of the rigidity parameter the phenomenon of phase amplification prevents liquid-liquid phase separation, similar to that in the DBC model, making this transition equivalent to a lambda transition. However, if interconversion of chiral isomers could be prevented by a sufficiently large kinetic barrier, the phase behavior would be similar to that in a SCB lattice with liquid-liquid phase separation at different compositions and the formation of a fluid-fluid interface.

\subsection{Dependence on the Coordination Number}~\label{Sec_CoordinationNumber}

To quantify deviations of the location of the transition temperatures and densities from MF behavior in the DBC model, we define the reduced temperature and density shifts of a transition point, \textit{tr} (where $\text{tr}=\mathrm{CP}$, $\lambda$EP, or TCP), as
\begin{align} 
\Delta \hat{T}_\text{tr} &\equiv 1-\frac{\hat{T}_\text{tr}}{\hat{T}_\text{tr}^{\rm MF}}\label{Eqn_redTempX}\\ 
\Delta \rho_\text{tr} &\equiv 1-\frac{\rho_\text{tr}}{\rho_\text{tr}^{\rm MF}}\label{Eqn_redRhoX}
\end{align}

\begin{figure}[t!]
    \centering
    \includegraphics[width=0.99\linewidth]{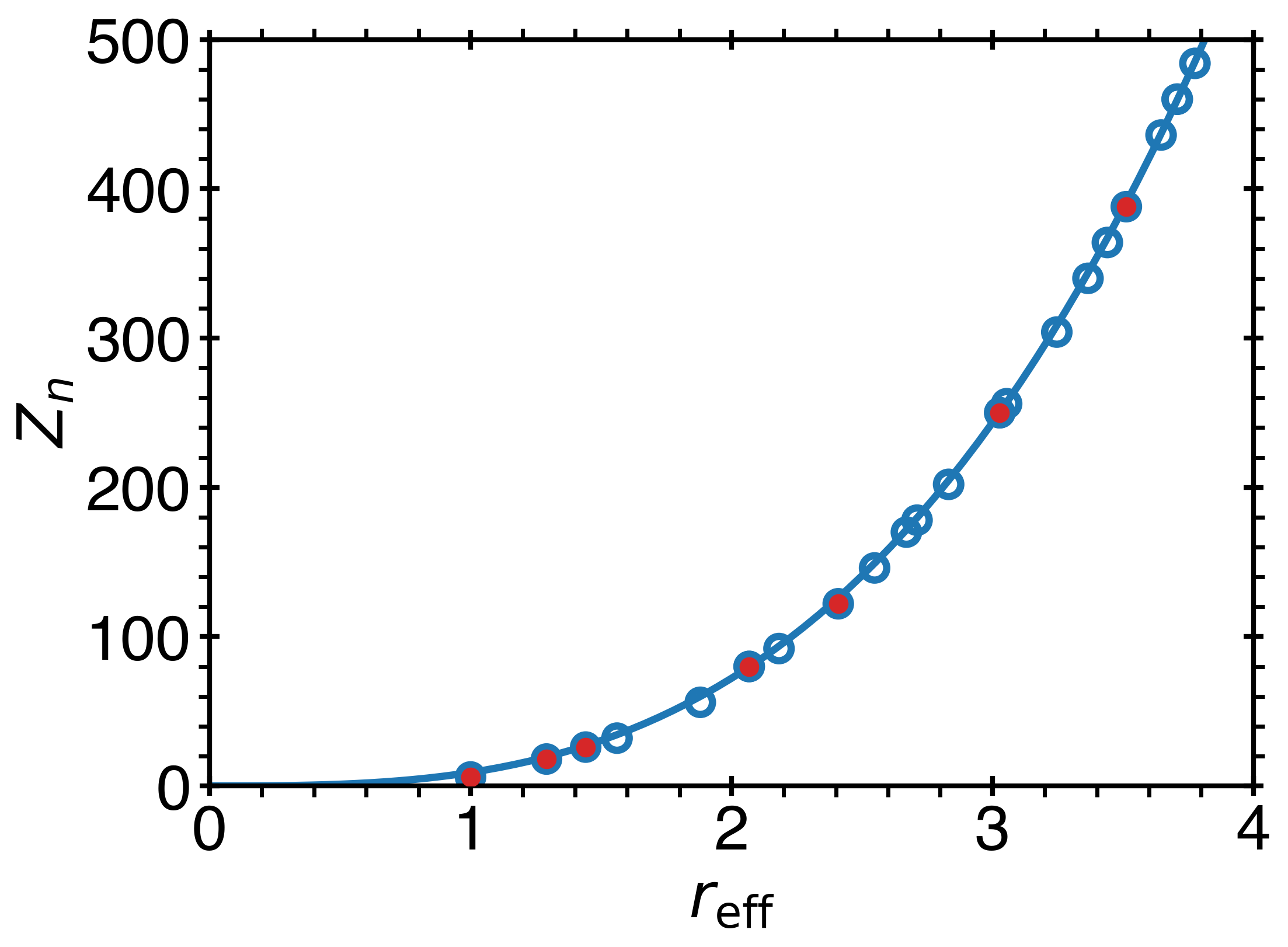}
    \caption{Dependence of the number of interacting neighbors, $Z_n$, on the effective interaction radius, $r_\text{eff}$. The circles show values calculated from Eq.~(\ref{Eq_r_eff}), while the solid curve shows the continuum-limit relation $Z_n= (4\pi/3)(5/3)^{3/2} {r_\text{eff}}^3\approx 9.013{r_\text{eff}}^3$, which follows from the volume of a sphere in the limit $Z_n\to\infty$. The circles with red centers correspond to the MC simulations considered in this work.}
    \label{Fig_Zn_vs_reff}
\end{figure}

\begin{figure*}
	\centering
	\includegraphics[width=0.32\linewidth]{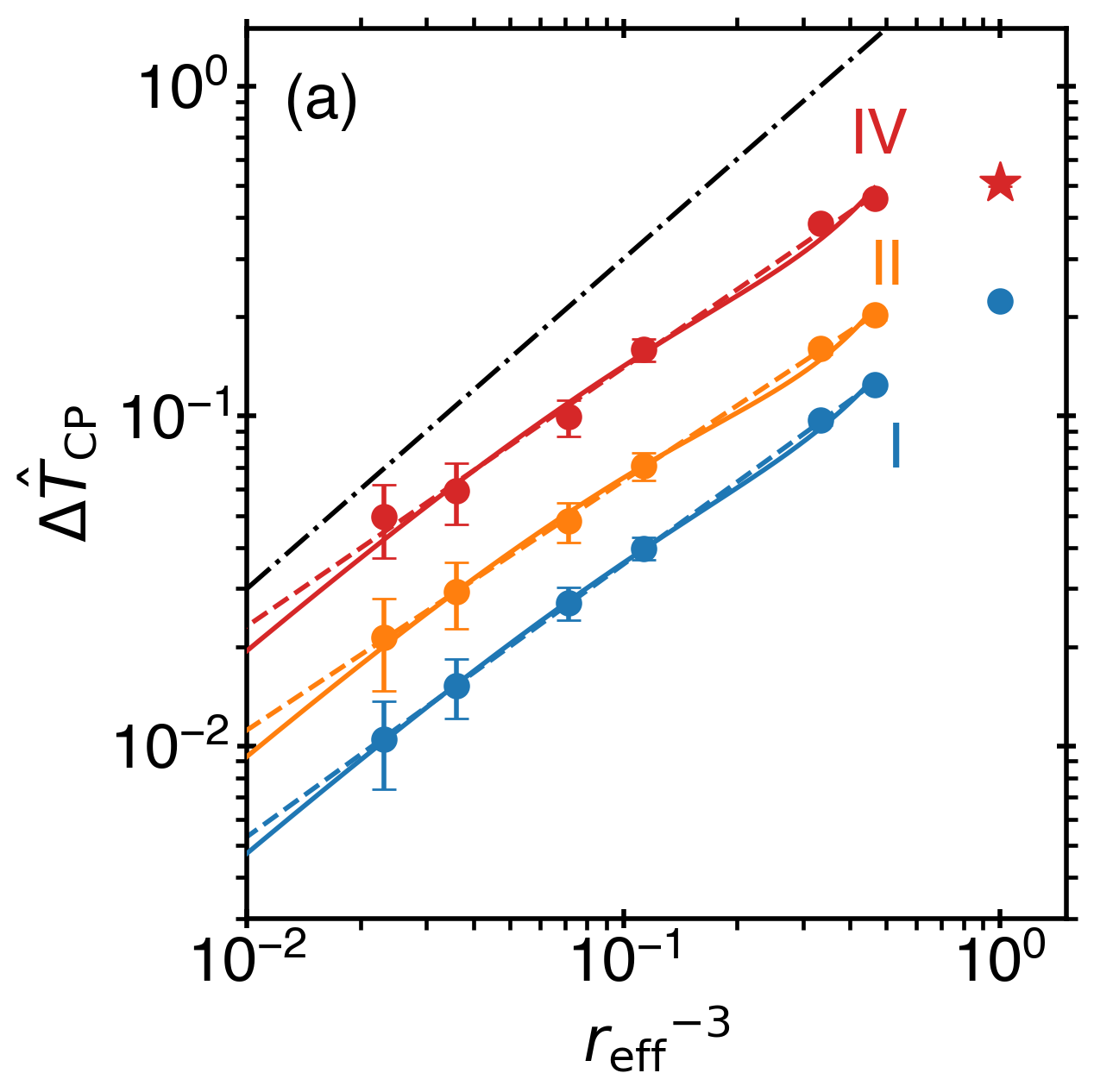}
	\includegraphics[width=0.32\linewidth]{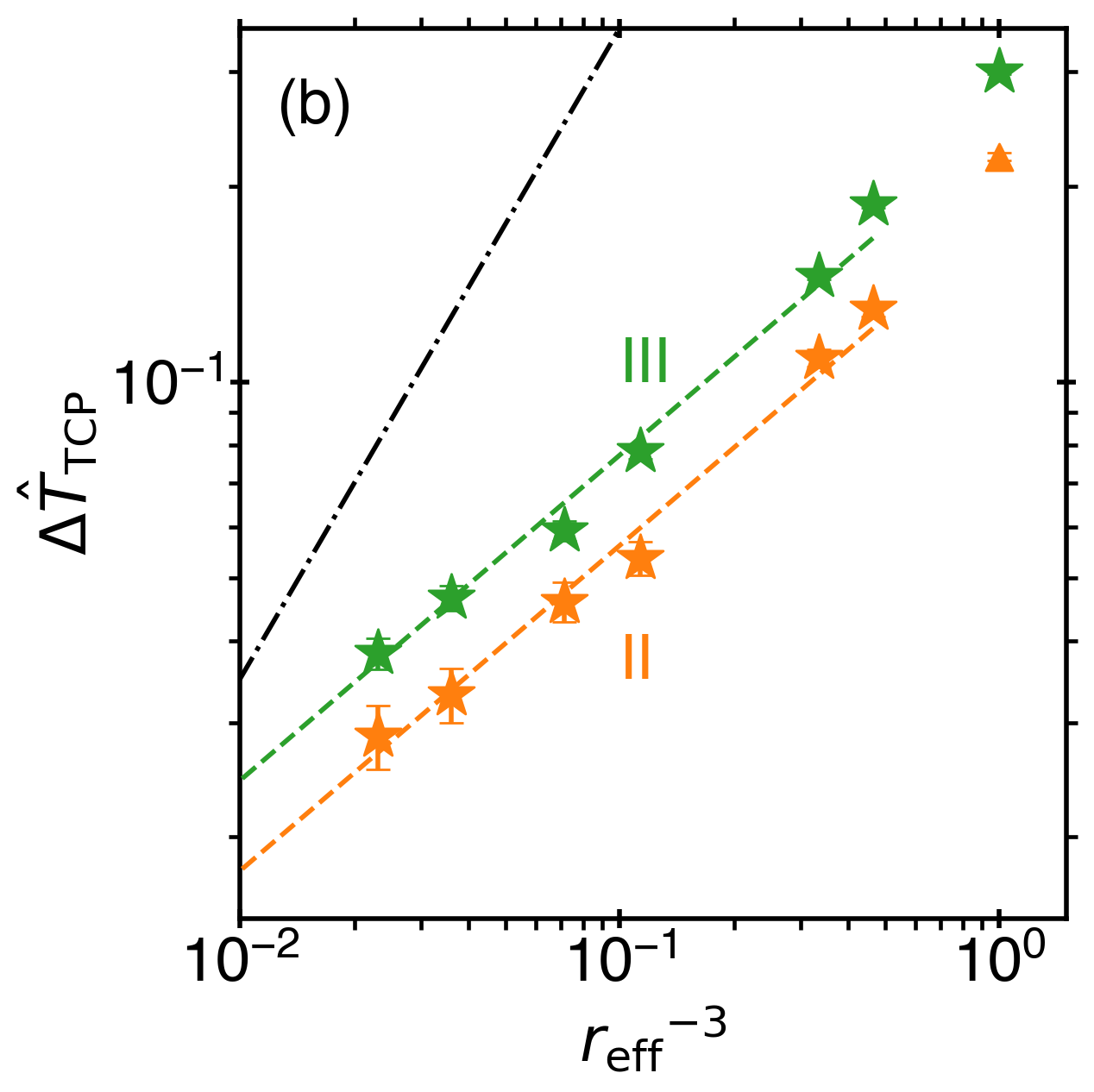}
	\includegraphics[width=0.32\linewidth]{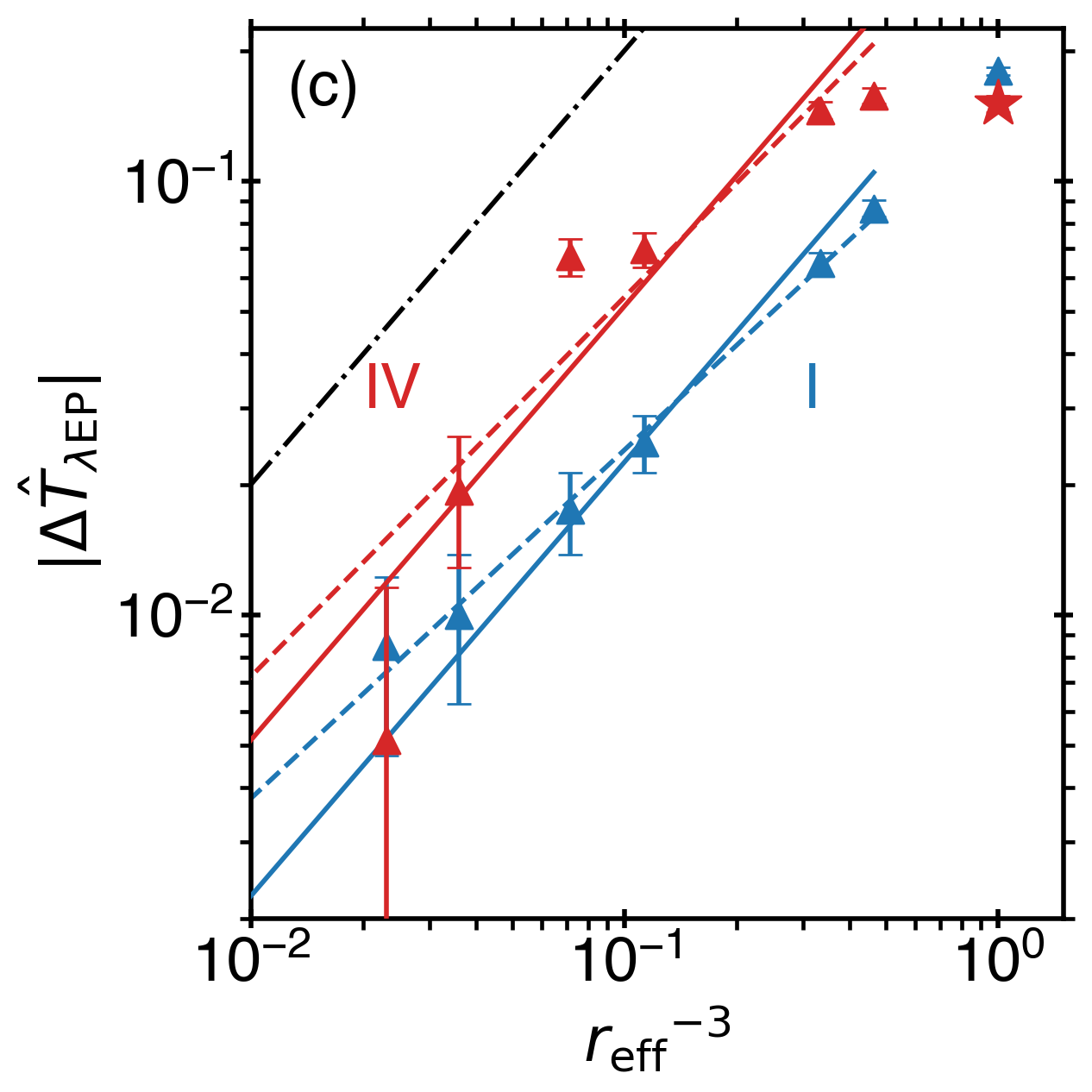}
	\includegraphics[width=0.32\linewidth]{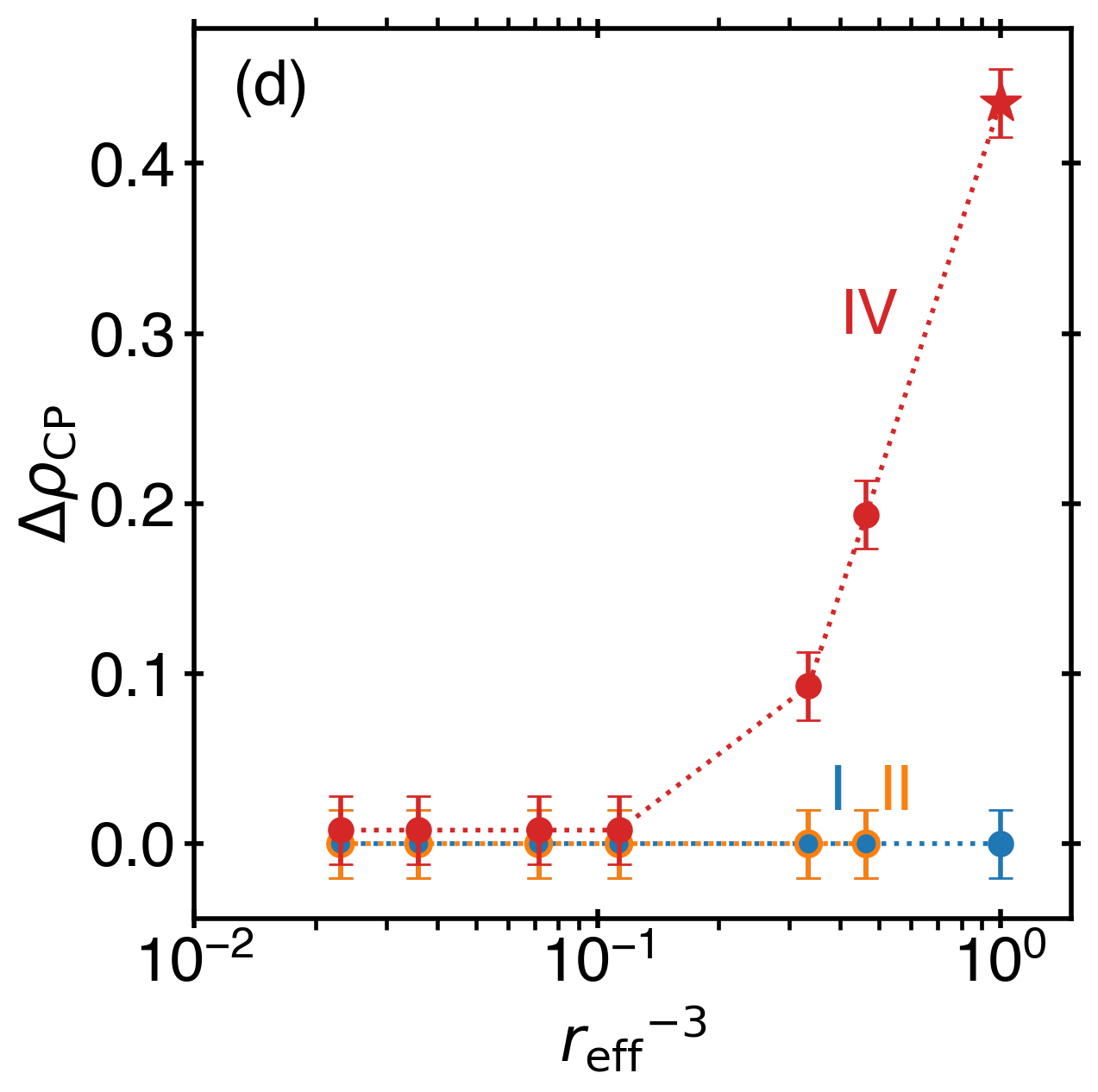}
	\includegraphics[width=0.32\linewidth]{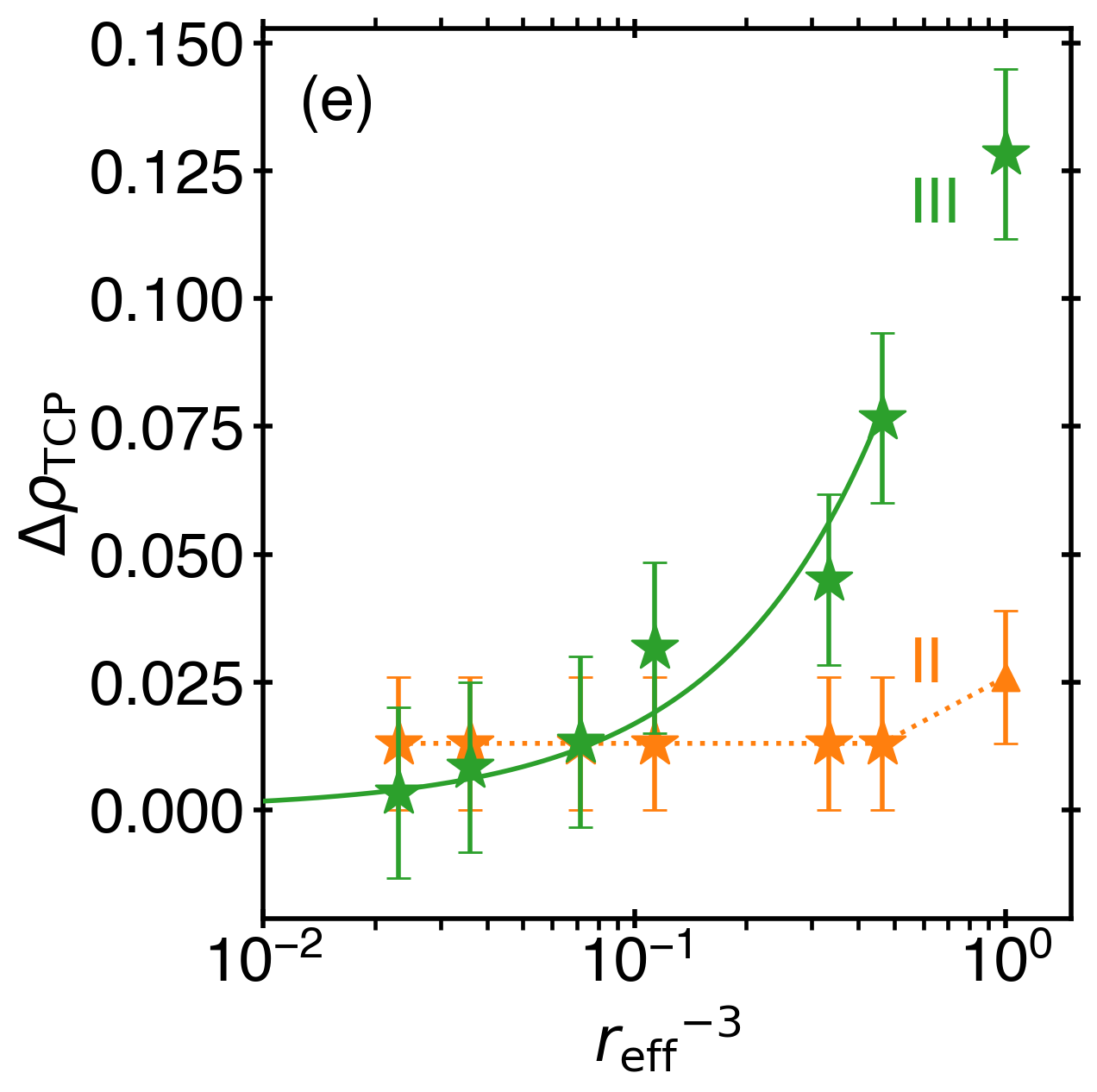}
	\includegraphics[width=0.32\linewidth]{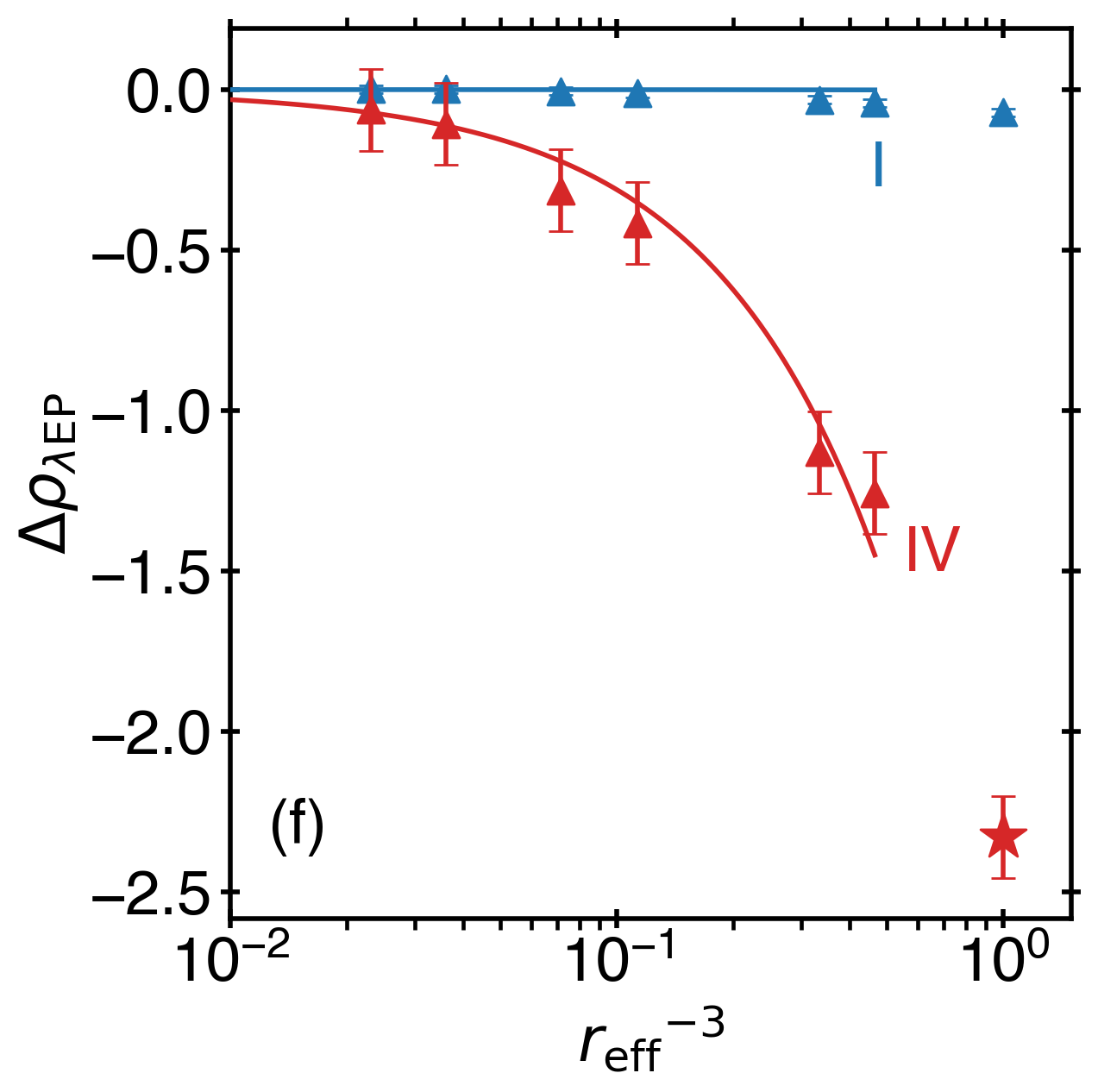}
\caption{Relative shifts of the transition temperatures (upper row) and densities (lower row) obtained from MC simulations with respect to their MF values, defined by Eq.~(\ref{Eqn_redTempX}) and  Eq.~(\ref{Eqn_redRhoX}), respectively, as functions of ${r_\text{eff}}^{-3}$. The panels correspond to the critical point (a,d), tricritical point (b,e), and $\lambda$-end point (c,f). Here, ${r_\text{eff}}^{-3}$ is proportional to $\sqrt{N_\text{G}}$ and $Z_n^{-1}$, where $N_\text{G}$ is the Ginzburg number and $Z_n$ is the number of interacting neighbors. The symbols denote MC results for the different archetypes: I (blue), II (orange), III (green), and IV (red). The solid curves represent fits of $\Delta\hat{T}_\text{CP}$ to Eq.~(\ref{eq:Tcp}) in (a), $\Delta\hat{T}_{\lambda{\text{EP}}}$ to $a\,{r_\text{eff}}^{-3}$ in (c), $\Delta\rho_\text{TCP}$ to $a\,{r_\text{eff}}^{-3}$ in (e), and $\Delta\rho_{\lambda\text{EP}}$ to $a\,{r_\text{eff}}^{-3}$ in (f). The power-law dependencies of $\Delta\hat{T}_{\lambda\text{EP}}$ and $\Delta\rho_{\lambda\text{EP}}$ are implied by Eqs.~(\ref{Eq_Geometric_DeltaT_LEP}) and (\ref{Eq_RhoShift_Comparison}), respectively, with prefactors absorbed into the fitting coefficient $a$. All fits were obtained by weighted least squares with relative weights ($\sigma=y$), excluding the ${r_\text{eff}}^{-3}=1$ ($Z_n=6$) data point. Dashed curves in panels (a-c) represent empirical power-law fits of the form $a\,{r_\text{eff}}^{-b}$, while the dashed-dotted lines illustrate the power-law ${r_\mathrm{eff}}^{-3}$. Dotted lines serve as guidelines for the eye. The fit parameters for $\Delta\hat{T}_\text{tr}$ are reported in Table~\ref{tab:deviation_fits}, while the fitted coefficients for $\Delta\rho_\text{tr}$ are $a=0.17$ (e), $-0.002$ for Type I in (f), and $-3.12$ for Type IV in (f). For clarity, the plotted values for Types II and IV in panel (a) are multiplied by factors of 2 and 5, respectively.}
	\label{Fig_CP_vs_ZN}
\end{figure*}
{\noindent}where $\hat{T}_\text{tr}^{\rm MF}$ and $\rho_\text{tr}^{\rm MF}$ denote the corresponding MF transition values. The MF coordinates of the critical, $\lambda$-end, tricritical points for all four archetypes are provided in Table~\ref{tab_MF_critVals}.

MC simulations of the DBC model were conducted for a range of coordination numbers from $Z_n=6$ to $Z_n=388$, for four values of $\bar{\omega}$, representing the four archetypal regions. The temperature and density coordinates of the CP, TCP, and/or $\lambda$EP are presented in Fig.~\ref{Fig_CP_vs_ZN} as functions of the ``effective'' range of interactions, ${r_\text{eff}}$,~\cite{Luijten_Critical_1999} related to the coordination number via 
\begin{equation}\label{Eq_r_eff}
    {r_\text{eff}}^2 = \frac{1}{Z_n}\sum_{i\neq j}|r_i -r_j|^2 \quad\text{where}\quad |r_i-r_j|\le r_z
\end{equation}
which goes to ${r_\text{eff}}\to \sqrt{3/5}r_z$ in the continuum limit ($Z_n\to\infty$). As shown in Fig.~\ref{Fig_Zn_vs_reff}, the number of interacting neighbors scales approximately as $Z_n \propto {r_\mathrm{eff}}^3$ with a proportionality coefficient $(4\pi/3)(5/3)^{3/2} \approx 9.013$, which follows from the volume of a sphere in the continuum limit. As noted by Luijten~\cite{Luijten_Critical_1999}, finite-size effects on the location of the transition point are expected to be negligible when the system size satisfies $\ell \gtrsim {r_\text{eff}}^4$. In the present work, $\ell=2^{1/3}\ell_x\approx161$, while the largest interaction range considered was ${r_\text{eff}}\approx3.51$, corresponding to ${r_\text{eff}}^4\approx152$. Thus, the systems with the largest ${r_\text{eff}}$ values satisfy this criterion only marginally, indicating that finite-size effects may begin to influence the transition-point locations for these systems.

The error bars shown in Fig.~\ref{Fig_CP_vs_ZN} were estimated by dividing the MC uncertainty by the MF value of the corresponding transition point. All fits are weighted least squares fits ($\sigma=y$) and exclude the $Z_n=6$ data point (${r_\text{eff}}^{-3}=1$) as lattice effects become significant for the shortest interaction range (see Section~\ref{Sec_PhaseDiagrams}). In this subsection, we examine the convergence of the thermodynamic-point coordinates toward their corresponding MF predictions as the coordination number increases.

\paragraph{Critical Point.} The convergence of the locations of the critical-points toward their MF values may be understood within a crossover theory of criticality developed for the $3d$ Ising universality class, which relates the size of the critical region through the Ginzburg criterion~\cite{Luijten_Nature_1998,kim_crossover_2003,Luijten_Critical_1999,Binder_Crossover_2001}. Using renormalization group (RG) analysis~\cite{Luijten_Nature_1998,Luijten_Critical_1999,Binder_Crossover_2001}, various critical thermodynamic properties of Ising-like system may be expressed in terms of the effective size of the critical region through the Ginzburg number ($N_\text{G}$), which depends on ${r_\text{eff}}$, given by Eq.~(\ref{Eq_r_eff}), as
\begin{equation}\label{Eq_Ginz_CP}
    N_\text{G} = N_\text{G}^\circ \left({r_\text{eff}}\right)^{-d/\phi}
\end{equation}
where $N_\text{G}^\circ$ is a constant and $\phi = (d_\text{m}-d)/2$ in which, for Ising-like systems, the marginal dimensionality, $d_\text{m}$, is $d_\text{m}=4$~\cite{Goldenfeld_Lectures_2018,Wilson_Critical_1972,Wilson_RG_1975}. The crossover from the fluctuation-affected critical region to the MF critical region is established through the Ginzburg criterion~\cite{Amit_Ginzburg_1974, Fisher_RG_1974,Sengers_Critical_2009}. Thus, in $d=3$, with use of Eq.~(\ref{Eq_Ginz_CP}), one finds that $N_\text{G}(d\to3)/N_\text{G}^\circ = {r_\text{eff}}^{-6}$ and since $N_\text{G}\propto\Delta\hat{T}$, the crossover occurs when $\Delta\hat{T}\ll N_\text{G}^\circ {r_\text{eff}}^{-6}$ or when $z=\Delta\hat{T}{r_\text{eff}}^6/N_\text{G}^\circ\sim 1$.

The shift in the critical temperature with respect to its MF value has been predicted~\cite{Brout_Ferromagnetism_1960,Thouless_Ising_1969,Luijten_Nature_1998,Luijten_Critical_1999,Binder_Crossover_2001,Grazieschi_IsingKac_2025} to scale with the Ginzburg number according to $\Delta\hat{T}_\text{CP}\sim N_\text{G}^\phi\sim {r_\text{eff}}^{-3}$. This prediction is approximately obeyed by our results (Fig.~\ref{Fig_CP_vs_ZN}a), but some deviation is apparent. Several corrections to the leading ${{r_\text{eff}}}^{-3}$ have been proposed in the literature. Thouless~\cite{Thouless_Ising_1969} introduced a subleading term proportional to $(\ln {r_\text{eff}})/{{r_\text{eff}}}^{6}$. Luijten~\cite{Luijten_Nature_1998} subsequently argued that RG analysis generates an additional contribution proportional to ${{r_\text{eff}}}^{-5}$ arising from the renormalization of both the quadratic and gradient terms in the Landau-Ginzburg-Wilson Hamiltonian, leading to:
\begin{equation}\label{eq:Tcp_old}
    \Delta\hat{T}_\text{CP} =  \frac{a}{{r_\text{eff}}^3} + \frac{b}{{r_\text{eff}}^5} + \frac{c\ln{{r_\text{eff}}} + d}{{r_\text{eff}}^6}
\end{equation}
where $a$, $b$, $c$, and $d$ are constants. More recently, Grazieschi et al.~\cite{Grazieschi_IsingKac_2025} adopted a description of the Ising-Kac model containing only the leading ${r_\text{eff}}^{-3}$ term and the logarithmic correction:
\begin{equation}\label{eq:Tcp}
    \Delta\hat{T}_\text{CP} = \frac{a}{{r_\text{eff}}^3} + \frac{b \ln {r_\text{eff}}}{{r_\text{eff}}^{6}} \, .
\end{equation}
Due to the limited number of simulation data points, which span approximately only one decade in $\Delta\hat{T}_{\text{CP}}$, and since the largest interaction range considered ($Z_n=388$) may be affected by residual finite-size effects, the present data do not provide sufficient resolution to reliably determine the form of the correction to the ${r_\text{eff}}^{-3}$ asymptotic behavior. Therefore, we selected Eq.~(\ref{eq:Tcp}) which contains only two free parameters and fits the data well. We note however that the simulation results are also well described by an empirical power law $a\,{r_\text{eff}}^{-b}$, with a best-fit exponent $b\approx 2.3$. The corresponding fit parameters are reported in Table~\ref{tab:deviation_fits}. Within the uncertainty of the simulation data and the limited range of interaction lengths explored, the present results do not allow a clear distinction between the empirical power-law description and the theoretically motivated crossover form (Eq.~\ref{eq:Tcp}). Larger interaction ranges (simulated in larger systems) may be required to resolve the asymptotic regime predicted by the crossover theory.

The corresponding critical-density behavior is shown in Fig.~\ref{Fig_CP_vs_ZN}d. We observe no systematic dependence of the reduced critical density, $\Delta\rho_{\rm CP}$, on coordination number for the Type I and Type II archetypes. This result is in agreement with the theory of correspondence between the Ising model and the lattice-gas model and is a specific feature of symmetric systems~\cite{Lee_LatticeGas_1952,Wilding_asymmetric_1993}. Consequently, the critical density appears to be effectively independent of the interaction range for archetypes I and II. However, for archetype IV (emerging only for $Z_n>6$) the critical density depends on the interaction range, since the liquid-gas critical line intercepts the tricritical line (archetype III) at $\bar{\omega}\approx 2.5$ as seen in Fig.~\ref{fig_Zn6_symmetry}.

\paragraph{Tricritical Point.} The behavior of the tricritical-point fundamentally differs from that of a conventional critical point because the marginal dimensionality for tricriticality is $d_\text{m}=3$,~\cite{Riedel_Tricritical_1972} such that $\phi = (3-d)/2$. Consequently, for a $d=3$ system, $\phi=0$ and the Ginzburg number is undefined. Hence, the conventional Ginzburg-number-based scaling used for critical points does not apply~\cite{Lawrie_Tricritical_1984}. Nevertheless, RG theories of tricriticality predict logarithmic corrections to the MF behavior due to tricritical fluctuations~\cite{Lawrie_Tricritical_1984,Riedel_Tricritical_1972,Wegner_Logarithmic_1973,Fisher_RG_1975,Stephen_Logarithmic_1975}. These logarithmic corrections have been derived for a variety of tricritical quantities, including finite-size scaling relations and thermodynamic scaling laws near the tricritical point~\cite{Lawrie_Tricritical_1984,Moueddene_Blume_2024,Li_Logarithmic_2024}. However, to the best of our knowledge, unlike the critical-point case, a corresponding RG prediction for the dependence of the tricritical-temperature shift on the effective interaction range has not been established. Consequently, while logarithmic corrections are expected on general theoretical grounds, the functional form governing the crossover of $\Delta\hat{T}_{\text{TCP}}$ with increasing ${r_\text{eff}}$ remains uncertain.

\begin{table}[t]
\caption{\label{tab:deviation_fits}
Fit parameters describing the deviations of the transition temperatures obtained from MC simulations to their MF values. (a) Parameters obtained by fitting  $\Delta\hat{T}_\text{CP}$ to Eq.~(\ref{eq:Tcp}) and $\Delta\hat{T}_{\lambda\text{EP}}$ to $a\,{r_\text{eff}}^{-3}$, as follows from Eq.~(\ref{Eq_Geometric_DeltaT_LEP}). (b) Parameters obtained by fitting each $\Delta\hat{T}_\text{tr}$ to an empirical power-law of the form $a\,{r_\text{eff}}^{-b}$.}
\begin{ruledtabular}
\begin{tabular}{c c cc cc cc}
& 
& \multicolumn{2}{c}{CP}
& \multicolumn{2}{c}{$\lambda$EP}
& \multicolumn{2}{c}{TCP}\\
$\bar{\omega}$
& Archetype
& $a$ & $b$
& $a$ & $b$
& $a$ & $b$ \\
\hline

\multicolumn{8}{c}{\textit{(a) theory (solid curves on Fig.~\ref{Fig_CP_vs_ZN})}} \\
0.4 & I   & 0.50 &   -1.85 & 0.23   & --  & --   & --   \\
0.5 & II  & 0.50 &   -2.27 & --     & --  & --   & -- \\
1.0 & III & --   & --      & --     & --  & --   & -- \\
5.0 & IV  & 0.41 &   -1.71 & -0.52  & --  & --   & --   \\
\hline
\multicolumn{8}{c}{\textit{(b) Empirical (dashed lines on Fig.~\ref{Fig_CP_vs_ZN})}} \\
0.4 & I   & 0.24 & 2.48 & 0.15   & 2.41  & --     & --  \\
0.5 & II  & 0.18 & 2.27 & --     & --    & 0.18   & 1.50 \\
1.0 & III & --   & --   & --     & --    & 0.24   & 1.50 \\
5.0 & IV  & 0.17 & 2.35 & -0.41  & 2.62  & --     & --  \\
\end{tabular}
\end{ruledtabular}
\end{table}

Figure~\ref{Fig_CP_vs_ZN}b shows that the reduced tricritical-temperature shift, $\Delta\hat{T}_{\text{TCP}}$, varies much more gradually than the ${r_\text{eff}}^{-3}$ asymptotic behavior predicted for conventional criticality, $\Delta\hat{T}_{\text{CP}}$. Over the range investigated, the simulation results are reasonably described by an empirical power law of the form $a\,{r_\text{eff}}^{-3/2}$. The corresponding fit parameters are given in Table~\ref{tab:deviation_fits}. Given the uncertainty of the simulation data, the limited variation in $\Delta\hat{T}_{\text{TCP}}$ (approximately one decade), and the possibility of finite-size effects at the largest ${r_\text{eff}}$ values investigated, the present results do not allow reliable discrimination between alternative crossover forms. Therefore, no attempt is made to infer the functional dependence of $\Delta\hat{T}_{\text{TCP}}$ on ${r_\text{eff}}$ from the available data. Nevertheless, the comparatively slow approach toward the MF limit with increasing interaction range is qualitatively consistent with the expectation that logarithmic corrections play an important role in tricritical crossover phenomena at the marginal dimensionality~\cite{Wegner_Logarithmic_1973,Fisher_RG_1975,Stephen_Logarithmic_1975}.

Further support for the behavior of $\Delta\hat{T}_\text{TCP}$ is given in the following studies. In a vectorized BEG model, Maciołek \textit{et al.}~\cite{Maciolek_helium_2004}, found that the reduced tricritical temperature is independent of coordination number within molecular-field theory, whereas their MC simulations of a simple-cubic lattice ($Z_n=6$) exhibited behavior consistent with logarithmic corrections. We note that because only a single coordination number was examined numerically the study did not directly address the dependence of the tricritical coordinates on ${r_\text{eff}}$~\cite{Maciolek_helium_2004}. Sch\"oll-Paschinger \textit{et al.}~\cite{Paschinger_binary_2005} investigated the influence of interaction range on symmetric binary mixtures with a HCY potential, demonstrating qualitative changes in tricritical coordinates and phase behavior using MC simulations.

Figure~\ref{Fig_CP_vs_ZN}e shows that the behavior of the reduced tricritical density, $\Delta\rho_{\text{TCP}}$, is not uniform across the archetypes. A clear dependence on ${r_\mathrm{eff}}$ is observed for the Type III system ($\bar{\omega}=1$), whereas no statistically significant dependence is resolved for the Type II system ($\bar{\omega}=0.5$). These observations do not support a universal interaction-range dependence of the tricritical density. Instead, any residual variation of $\Delta\rho_{\text{TCP}}$ appears to be system dependent. Given the uncertainty of the MC density estimates ($\delta\rho\approx\pm0.01$), the available data are insufficient to establish either a universal functional dependence or the presence of exclusively logarithmic corrections.

\begin{figure}[tpb]
    \centering
    \includegraphics[width=0.95\linewidth]{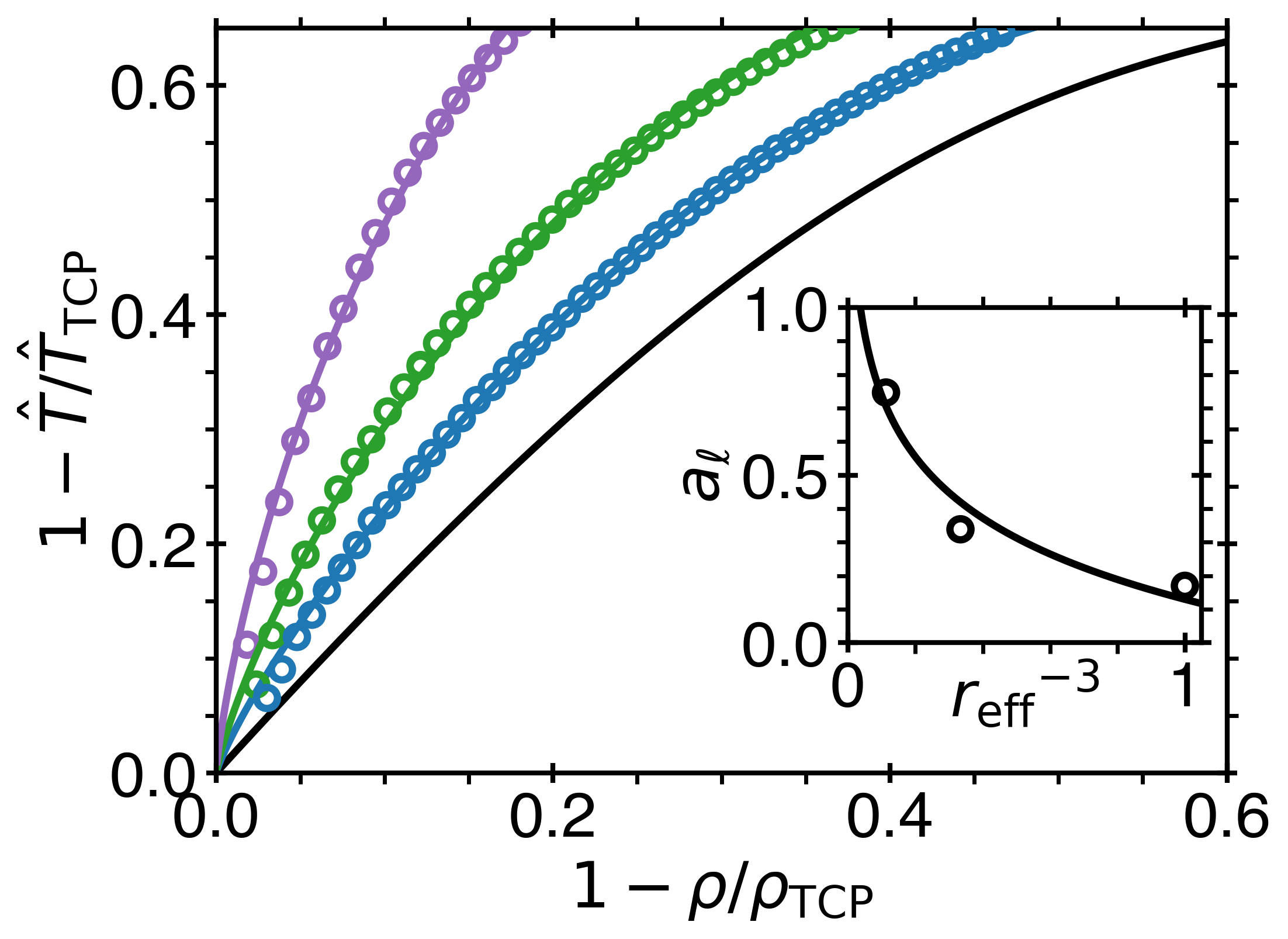}
    \caption{Reduced liquid branch of coexistence for the system with $\bar{\omega}=1$. Colors indicate the three $Z_n$ values considered in Section~\ref{Sec_PhaseDiagrams}: $Z_n=6$ (purple), $Z_n=26$ (green), and $Z_n=80$ (blue). The black curve denotes the MF theory prediction. Open circles correspond to MC simulation data, while the colored curves are fits to Eq.~(\ref{Eq_LiquidBranch_Tricrit}). The fitted exponent was found to be $b_\ell=1.22$ for all three systems, whereas the amplitude parameter varies with $Z_n$ as: $a_\ell=0.170$ ($Z_n=6$), $0.338$ ($Z_n=26$), and $0.746$ ($Z_n=80$). The inset shows the dependence of the fitted amplitude parameter $a_\ell$ on ${r_\text{eff}}^{-3}$. Open circles denote the fitted $a_\ell$ values, and the solid curve represents the empirical logarithmic fit, $a_\ell =-0.264 \ln({r_\text{eff}}^3)+0.129$.}
    \label{Fig_tricrit_amplitude}
\end{figure}

The effect of logarithmic corrections is also present in the amplitude of the liquid branch of coexistence. This dependence was examined for the system with $\bar{\omega}=1$, which exhibits the clearest signature of tricriticality among the four archetypes investigated. RG theory predicts the scaling relation~\cite{Lawrie_Tricritical_1984,Fisher_RG_1975,Moueddene_Blume_2024},
\begin{equation}
\varphi\sim \tau_\text{TCP}^\beta|\ln\tau_\text{TCP}|^{\hat{\beta}}
\end{equation}
in the asymptotic vicinity of the fluctuation-affected TCP where $\varphi=1-\rho/\rho_\text{TCP}$ is the order parameter, $\tau_\text{TCP} = 1 - \hat{T}/\hat{T}_\text{TCP}$ is the reduced distance to the tricritical point, and $\hat{\beta}$ is the logarithmic correction to the asymptotic critical exponent ($\beta=1$ for $3d$ tricritical systems). Consistent with this relation, the liquid branch was found to follow
\begin{equation}\label{Eq_LiquidBranch_Tricrit}
    \varphi = \varphi^\text{MF}\left[1 + a_\ell \left|\ln\tau_\text{TCP}\right|^{b_\ell}\right]
\end{equation}
where $\varphi^\text{MF}=1-\rho^\text{MF}/\rho_\text{TCP}^\text{MF}$ is the order parameter of the MF liquid branch of the coexistence curve and is $\varphi^\text{MF}\propto\tau_\text{TCP}$ in the vicinity of the TCP, while $a_\ell=a_\ell(Z_n)$ and $b_\ell$ are constants. Figure~\ref{Fig_tricrit_amplitude} compares this expression with MC simulation results for the three $Z_n$ values considered in Section~\ref{Sec_PhaseDiagrams} ($Z_n=6$, $26$, and $80$), demonstrating a strong agreement with Eq.~(\ref{Eq_LiquidBranch_Tricrit}).

\paragraph{Lambda End Point.} To the best of our knowledge, no theoretical description has been established for the dependence of the $\lambda$EP temperature or density on ${r_\text{eff}}$. Since the $\lambda$EP is the intersection of the second-order $\lambda$-transition line with the liquid-vapor coexistence curve, its location is determined jointly by the position of these two phase boundaries. In the DBC model, these boundaries also depend on $\bar{\omega}$, but at fixed $\bar{\omega}$, the dominant dependence is governed by fluctuation-induced corrections proportional to ${r_\text{eff}}^{-3}$. Since the $\lambda$EP and the CP both belong to the $3d$ Ising-model universality class, it is reasonable to expect that their leading temperature shifts are governed by the same Ginzburg-scaling arguments, such that $\Delta\hat{T}_{\lambda\text{EP}}\sim\Delta\hat{T}_\text{CP}\sim N_\text{G}^\phi\sim {r_\text{eff}}^{-3}$. Additionally, because the $\lambda$EP is constrained to lie on the coexistence curve, the associated shift in density is hypothesized to inherit the same leading-order scaling behavior, namely that $\Delta\rho_{\lambda\text{EP}}\sim {r_\text{eff}}^{-3}$. Scaling arguments, using the geometry of the liquid-vapor coexistence and the $\lambda$-line, are presented in Appendix~\ref{Sec_Appendix_GeometricArguments} that support this hypothesis. This expectation may also be supported by the MC simulation results presented in Figs.~\ref{Fig_CP_vs_ZN}(c,f). In particular, $\Delta\hat{T}_{\lambda\text{EP}}$ is reasonably described by $a\,{r_\text{eff}}^{-3}$, which follows from Eq.~(\ref{Eq_Geometric_DeltaT_LEP}). However, an empirical power law of the form $a\,{r_\text{eff}}^{-b}$ provides an equally good representation of the data over the range investigated. The fit parameters for both curves are presented in Table~\ref{tab:deviation_fits}. For the $\lambda$EP density shift, only the leading-order behavior could be reliably resolved. 

As indicated by the coefficients in Table~\ref{tab:deviation_fits}, the simulation data show that $\Delta\rho_{\lambda\text{EP}}$ is negative for both Type I and Type IV behavior, while $\Delta\hat{T}_{\lambda\text{EP}}$ is positive for Type I and negative for Type IV. These sign differences are consistent with the geometric arguments presented in Appendix~\ref{Sec_Appendix_GeometricArguments}. In particular, the observed decrease of the $\lambda$-line slope relative to its MF value implies $a_\lambda\le0$, such that Eq.~(\ref{Eq_RhoShift_Comparison}) predicts that the density shift results from a competition between the displacement of the $\lambda$-line and the temperature shift of the end point. Like the CP and TCP, the uncertainty of the simulation data, the limited range of shift values examined, and the possibility that the largest interaction ranges are affected by finite-size effects or have not yet reached the asymptotic regime prevent a clear distinction between the empirical and theoretical descriptions. Consequently, the present results are consistent with, but do not provide definitive evidence for, the hypothesized behavior that $\Delta\hat{T}_{\lambda\text{EP}}\sim\Delta\rho_{\lambda\text{EP}}\sim\Delta\hat{T}_\text{CP}\sim {r_\text{eff}}^{-3}$.

\section{Conclusion}~\label{Sec_Conclusion}
In this work, we investigated the dependence of the range of molecular interactions, by varying the coordination number, on the multicritical phase behavior of a minimal microscopic model of a single-component substance, referred to as the ``degenerate blinking-checkers'' (DBC) model.~\cite{Caupin_Polyamorphism_2021,Buldyrev_BCM_2024,Anisimov_Degenerate_2025} Using Molecular dynamics simulations, we demonstrated that in the limit of an infinite range of interactions, the $3d$ MC simulations are in full agreement with the MF predictions. However, when the interaction range is finite, the simulations are strongly affected by critical fluctuations. While simulations of systems with larger interaction ranges, with the coordination number $Z_n>6$, demonstrate all four archetypes of multicritical behavior, like those found in the meanfield theory, for $Z_n=6$ only two types are observed: I and III.

We predict that the multicritical phase behavior observed in the DBC model may also be relevant to other systems with degenerate molecular interconversion, such as mixtures of interconverting chiral isomers~\cite{Latinwo_MolecModel_2016,Uralcan_Interconversion_2020,Petsev_Effect_2021,Wang_Chiral_2022}. To establish the extent of this correspondence would be desirable, but would also require more detailed simulations of the phase behavior in the chiral model, in a broad range of the bias parameter (analogous to $\bar{\omega}$ in the DBC model), and a direct comparison with the results of the DBC model.

\acknowledgments
SVB acknowledges the use of the Shared Computing Cluster managed by the Boston University Research Computing Services for the simulations. The authors also wish to thank Hajime Tanaka for inspiration and stimulating discussions throughout the years.

\section*{Data Availability Statement}
The data that support the findings of this study are available from the corresponding author upon reasonable request.

\appendix

\section{Nearest-Neighbor DBC Model at $\bar{\omega}=0.5$}\label{Sec_Appendix_SpecialCase}

In this appendix, we examine the unique phase behavior of the nearest-neighbor (NN) DBC model at $\bar{\omega}=0.5$, which exhibits a flat coexistence boundary in $T$-$\rho$ space (see Fig.~\ref{Fig_cxc}b). At this non-ideality value, the NN-DBC model can be mapped onto the standard three-state Potts model~\cite{Wu_Potts_1982} on a three-dimensional simple-cubic lattice, as described by the Hamiltonian
\begin{equation}
    H_\text{P} = -J_\text{P}\sum_{\langle i,j\rangle}\delta_{s_i,s_j}
\end{equation}
where $s_i=0,1,2$ denotes the state at lattice site $i$, $\delta_{s_i,s_j}$ is the Kronecker delta, and the sum runs over nearest-neighbor interacting pairs. The ferromagnetic coupling $J_\text{P}>0$ favors neighboring sites in the same state.

To establish the correspondence with the DBC model, we identify state $0$ with an empty site and states $1$ and $2$ with the two interconverting species. Both the DBC and three-state Potts models have six possible unique pair interactions, but the models assign each pair different interaction energies. In the DBC model, all interactions involving empty sites are zero, leaving only three nonzero interactions: $\omega_{11}$, $\omega_{22}$, and $\omega_{12}$. At $\bar{\omega}=0.5$, the interactions exhibit symmetry as $\omega_{11}=\omega_{22}=2\omega_{12}$. In contrast, the Potts model assigns equal interaction energies to pairs in the same state ($00$, $11$, and $22$), while interactions between different states are zero. Despite these differences, the two models are equivalent below the transition temperature, provided that $J_\text{P}=\omega_{11}/Z_n$. Thus, the NN-DBC model with $Z_n=6$ and $\omega_{11}=2$ coincides with the standard three-state Potts model with $J_\text{P}=1/3$.

The flat coexistence boundary can be understood from the symmetry of the three-state Potts model, which exhibits a first-order transition between disordered and ordered phases in three dimensions~\cite{Fukugita_Potts_1990,Villanova_Potts_1997,Wang_Potts_2014}. Denoting the fractions of sites in states $0$, $1$, and $2$ by $x_0$, $x_1$, and $x_2$, respectively, the three spin states are equally populated above the transition temperature, such that $x_0=x_1=x_2=1/3$. Below the transition temperature, an ordered phase emerges with a greater concentration of one spin state (e.g., $x_0$), while the other two remain equally populated, $x_{1}=x_2=(1-x_0)/2$. Due to the symmetry of the Hamiltonian, three equivalent ordered phases exist, each characterized by a greater concentration of one of the three spin states. To map this with the NN-DBC model, we denote the fraction of empty sites by $x_0$, such that the corresponding vapor density is $\rho_\text{G}=1-x_0$, while the liquid density is $\rho_\text{L}=1-(1-x_0)/2$. Thus, the two coexistence branches are symmetrically related by $\rho_\text{L}(T)=1-\rho_\text{G}(T)/2$. At or below the transition temperature, for a fixed overall density between $\rho_\text{G}(T)$ and $\rho_\text{L}(T)$, the system separates into two coexisting phases occupying different volumes, thereby conserving the total number of empty sites. At the Potts transition temperature, the discontinuity between the equilibrium densities gives rise to the flat portion of the NN-DBC coexistence boundary.

The phase diagram obtained from MC simulations shown in Fig.~\ref{Fig_cxc}(b) exhibits the principal features expected from the correspondence with the three-state Potts model. The estimated temperature of the flat coexistence boundary, $\hat{T}\approx0.6053$, agrees closely with the Potts transition temperature, $\hat{T}_\text{P}=0.60543866$.~\cite{Villanova_Potts_1997}, and the liquid and gas branches are consistent with the symmetry relation predicted from the symmetric Potts model. Defining the low-density and high-density endpoints of the interval bounding the discontinuous first-order transition as $\rho_\text{LD}$ and $\rho_\text{HD}$, the estimated endpoint densities of the NN-DBC model, $\rho_\text{LD}\approx 0.4$ and $\rho_\text{HD}\approx 0.8$, within the error bars, coincides with the discontinuity inferred from the Potts simulations of Shun \textit{et al.}~\cite{Wang_Potts_2014}, being approximately $\rho_\text{LD}=1-(1/3+0.2903)=0.376$ and $\rho_\text{HD}=0.812$, respectively. Overall, the mapping to the $3d$ three-state Potts model provides a physical interpretation of the unusual phase behavior observed at $\bar{\omega}=0.5$.

The equivalence to the standard three-state Potts model with $J_\text{P}=\omega_{11}/Z_n$ applies specifically to the NN-DBC model where $Z_n=6$. In this mapping, an energy shift induces an effective attraction between neighboring empty sites, represented as zero spins in the Potts model. For $Z_n>6$, however, the Potts model also introduces interactions between more distant zero spins, whereas the corresponding empty sites in the DBC model remain non-interacting. This distinction is reflected in the phase behavior shown in Fig.~\ref{Fig_cxc}(b), where increasing $Z_n$ leads to the emergence of distinct CP and TCP loci, associated with the low-density and high-density features of the phase diagram, respectively. Consequently, systems with $Z_n>6$ exhibit Type II behavior at $\bar{\omega}=0.5$, whereas the NN-DBC model exhibits a direct transition between Types I and III, with $\bar{\omega}=0.5$ representing a unique symmetric case. Thus, the correspondence with the three-state Potts model suggests that the absence of Type II behavior in the NN-DBC model is closely connected to the interplay between the underlying symmetry and the restriction to nearest-neighbor interactions.

\section{$\lambda$EP and $\lambda$-line Dependence on Interaction Range}\label{Sec_Appendix_GeometricArguments}

In this appendix, we provide a simple geometric argument showing that the reduced displacements of the $\lambda$-end point ($\lambda$EP) satisfy, $\Delta\hat{T}_{\lambda\text{EP}}\sim\Delta\rho_{\lambda\text{EP}}\sim\Delta\hat{T}_\text{CP}\sim {r_\text{eff}}^{-3}$, where $\Delta\hat{T}_{\text{tr}}$ and  $\Delta\rho_{\text{tr}}$, in which $\text{tr}=\text{CP}$ or $\lambda\text{EP}$, are given by Eq.~(\ref{Eqn_redTempX}) and (\ref{Eqn_redRhoX}) in the main text.

\begin{figure}[tpb]
    \centering
    \includegraphics[width=0.9\linewidth]{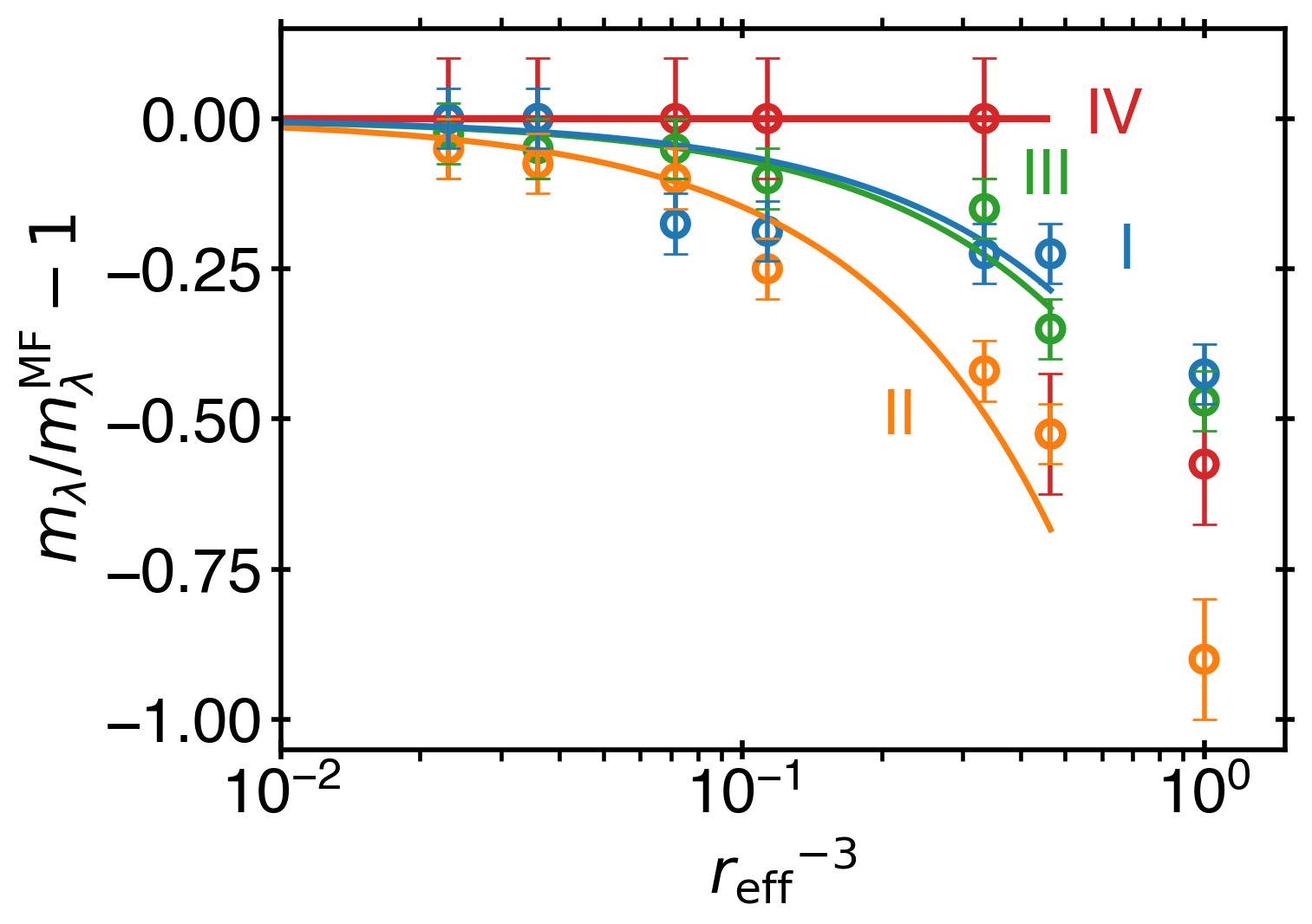}
    \caption{Reduced slope of the $\lambda$-transition line, $m_\lambda$, in the vicinity of the $\lambda$EP or TCP (if applicable) by its MF value $m_\lambda^\text{MF}=2\bar{\omega}$ as a function of ${r_\text{eff}}^{-3}$. The colors are different archetypes: $\bar{\omega}=0.4$, Type I, blue; $\bar{\omega}=0.5$, Type II, orange, $\bar{\omega}=1.0$, Type III, green; and $\bar{\omega}=5.0$, Type IV, red. The open circles are obtained from MC simulations with errors of $\pm 0.05$ (Types I, II, and III) and $\pm 0.1$ (Type IV), while the curves represent fits of Eq.~(\ref{Eq_app_aLambda}), in which $a_\lambda(\bar{\omega}=0.4)=-0.62$, $a_\lambda(\bar{\omega}=0.5)=-1.47$, $a_\lambda(\bar{\omega}=1.0)=-0.68$, and $a_\lambda(\bar{\omega}=5.0)=0.0$.}
    \label{Fig_lambda_line_vs_Zn}
\end{figure}

Consider a linear deviation from the MF limit of the form $\hat{T}_{\lambda\text{EP}} = \hat{T}_{\lambda\text{EP}}^\text{MF}(1-\Delta\hat{T}_{\lambda\text{EP}})$. Recall that the MF $\lambda$-line is given by $\hat{T}_{\lambda\text{EP}}^\text{MF} = 2\bar{\omega}\rho_{\lambda\text{EP}}^\text{MF}$. The fluctuation affected $\lambda$-line is empirically described by 
\begin{equation}\label{Eqn_EmpiricalLambda_Line}
    \hat{T}_{\lambda\text{EP}} = 2\bar{\omega}\left(1 + \frac{a_\lambda}{{r_\text{eff}}^{3}}\right)\rho_{\lambda\text{EP}}
\end{equation}
in the vicinity of the $\lambda$EP, where $a_\lambda$ is a constant that depends on $\bar{\omega}$, such that the MF limit is recovered when ${r_\text{eff}} \gg 1$. Fig.~\ref{Fig_lambda_line_vs_Zn} illustrates the linear dependence of the slope of the $\lambda$-line near the $\lambda$EP on ${r_\text{eff}}^{-3}$. Like the previous fits, the ${r_\text{eff}}=1$ ($Z_n=6$) system was excluded due to the lattice effects present in that system. Further support for this finding can be found in the vectorized VBEG model~\cite{Maciolek_helium_2004}. Although Maciołek \textit{et al.} did not explicitly investigate the dependence of the $\lambda$-line on interaction range, they theoretically predicted a $Z_n$-dependent slope of the $\lambda$-line. To further characterize the effect of fluctuations on the shape of the $\lambda$-line, we also examined its slope, $m_\lambda$, at $\rho=1$. In contrast to the clear $Z_n$ dependence observed in the vicinity of the $\lambda$EP, the values of $m_\lambda(\rho=1)$ for all $\bar{\omega}$ and $Z_n$ are consistent with the MFT predictions within the accuracy of the simulations. These findings indicate that fluctuations primarily affect the $\lambda$-line near the $\lambda$EP, where the line becomes progressively steeper, whereas away from the end point, the $m_\lambda$ remains close to its MF value. As $Z_n$ increases, these fluctuation-induced deviations, including the curvature of the $\lambda$-line, diminish and the MF limit is recovered. 

Rewriting Eq.~(\ref{Eqn_EmpiricalLambda_Line}) in terms of $\rho_{\lambda\text{EP}}$ and expanding to the leading order in $\Delta\hat{T}_{\lambda\text{EP}}$ yields 
\begin{equation}\label{Eqn_Rho_LambdaLine}
    \rho_{\lambda\text{EP}} \approx \rho_{\lambda\text{EP}}^\text{MF}\left(1 - \frac{a_\lambda}{{r_\text{eff}}^3} + \Delta \hat{T}_{\lambda\text{EP}}\right)
\end{equation}
A semi-empirical crossover equation connecting the MF and fluctuation-affected  coexistence curves is given by~\cite{Anisimov_PolymerTricrit_2024} 
\begin{equation}\label{Eqn_Crossover_Cxc}
    \Delta \rho=\Delta \rho^\text{MF}(1+z^2)^{(1-2\beta)/4\nu}
\end{equation}
where $\Delta \rho = 1-\rho/\rho_\text{CP}$ and $\Delta \rho^\text{MF} = 1-\rho^\text{MF}/\rho_\text{CP}^\text{MF}$, while $\beta$ and $\nu$ are critical exponents. Note that in the DBC model, all quantities depend on $\bar{\omega}$. In Eq.~(\ref{Eqn_Crossover_Cxc}), note that $\Delta\rho$ denotes the reduced deviation from the critical density as opposed to the coexistence density difference employed previously. The crossover parameter $z\equiv f({r_\text{eff}})^\nu |\Delta \hat{T}|^{-\nu}$, where $f({r_\text{eff}})\sim {r_\text{eff}}^{-3/\phi}$, in which $\phi=(d_\text{m}-d)/2$, as defined in the main text, and $\Delta \hat{T} = 1-T/T_\text{CP}$.  Equation~(\ref{Eqn_Crossover_Cxc}) has two limits: in the MF limit ($z\ll 1$), $\Delta \rho=\Delta \rho^\text{MF}$, while in the fluctuation-affected limit $z \gg 1$, $\Delta \rho = \Delta \rho^\text{MF} f^{1/2-\beta}({r_\text{eff}}) |\Delta T|^{\beta-1/2}$. Therefore, at fixed $\bar{\omega}$, the density is generally a function of both temperature and interaction range, $\rho=\rho(T,{r_\text{eff}})$. 

Assuming that $z$ is small but not nonzero, the coexistence curve can be expanded about the MF $\lambda$EP as
\begin{equation}\label{Eqn_Rho_Cxc_Expansion}
    \rho(T,{r_\text{eff}})\approx \rho_{\lambda\text{EP}}^\text{MF} + m_\text{cxc}\Delta\hat{T}_{\lambda\text{EP}} + A\Delta \hat{T}_\text{CP}
\end{equation}
where $m_\text{cxc}=\hat T_{\lambda\mathrm{EP}}^{\mathrm{MF}}(\partial \rho/\partial T)|_{{r_\text{eff}},\hat{T}=\hat{T}_{\lambda\text{EP}},\rho=\rho_{\lambda\text{EP}}}$ is the local slope of the coexistence curve evaluated at the $\lambda$EP and $A$ is a proportionality constant accounting for the displacement between the fluctuation-affected and MF coexistence curves. Equating Eqs.~(\ref{Eqn_Rho_LambdaLine}) and (\ref{Eqn_Rho_Cxc_Expansion}), recalling that $\Delta\hat{T}_\text{CP}\sim {r_\text{eff}}^{-3}$ to the leading order in ${r_\text{eff}}$ as given by Eq.~(\ref{eq:Tcp}), gives
\begin{equation}\label{Eq_Geometric_DeltaT_LEP}
    \Delta \hat{T}_{\lambda\text{EP}} = \left(\frac{A/\rho_{\lambda\text{EP}}^\text{MF} + a_\lambda}{1-2\bar{\omega}m_{\text{cxc}}}\right)\frac{1}{{r_\text{eff}}^3}
\end{equation}
Provided that the $\lambda$-line and coexistence curves do not have the same slope $1-2\bar{\omega}m_\text{cxc}\neq 0$, everything in parenthesis in Eq.~(\ref{Eq_Geometric_DeltaT_LEP}) is constant, so at fixed $\bar{\omega}$, $\Delta \hat{T}_{\lambda\text{EP}}\sim {r_\text{eff}}^{-3}$. Finally, using Eq.~(\ref{Eqn_Rho_LambdaLine}), it can be shown that 
\begin{equation}\label{Eq_RhoShift_Comparison}
    \Delta \rho_{\lambda\text{EP}} = \frac{a_\lambda}{{r_\text{eff}}^3} + \Delta \hat{T}_{\lambda\text{EP}}\sim \frac{1}{{r_\text{eff}}^3}
\end{equation}
Thus, at fixed $\bar{\omega}$, the observed ${r_\text{eff}}^{-3}$ variation of both the $\lambda$EP temperature and density is consistent with the ${r_\text{eff}}^{-3}$ displacement of the LGCP temperature established in Eq.~(\ref{eq:Tcp}), together with a corresponding leading-order ${r_\text{eff}}^{-3}$ correction to the $\lambda$-line.

Additional insight into the sign of the density shift can be obtained from the slopes of the $\lambda$-lines presented for each archetype in Fig.~\ref{Fig_lambda_line_vs_Zn}. From Eq.~(\ref{Eqn_EmpiricalLambda_Line}), the slope of the $\lambda$-line in the vicinity of the end point is 
\begin{equation}\label{Eq_app_aLambda} 
    \frac{m_\lambda}{m_\lambda^{\mathrm{MF}}}-1 = \frac{a_\lambda}{{r_\mathrm{eff}}^3}
\end{equation} 
Since all data shown in Fig.~\ref{Fig_lambda_line_vs_Zn} satisfy $(m_\lambda/m_\lambda^\text{MF}-1)\le0$, it follows that $a_\lambda\le0$ for all archetypes investigated, with $a_\lambda\approx0$ only for Type IV behavior. With use of Eq.~(\ref{Eq_RhoShift_Comparison}), the negative contribution of $a_\lambda$ is partially offset by the positive $\Delta\hat{T}_{\lambda\text{EP}}$ observed for Type I behavior, resulting in a small, but negative, $\Delta\rho_{\lambda\text{EP}}$, which is consistent with the blue curve shown in Fig.~\ref{Fig_CP_vs_ZN}f. In contrast, for Type IV behavior, $a_\lambda\approx0$ while $\Delta\hat{T}_{\lambda\text{EP}}<0$, such that both contributions to Eq.~(\ref{Eq_RhoShift_Comparison}) favor a negative density shift. Consequently, the magnitude of $\Delta\rho_{\lambda\text{EP}}$ is substantially larger for Type IV behavior than for Type I behavior, consistent with the red curve shown in Fig.~\ref{Fig_CP_vs_ZN}f. Thus, both the sign and relative magnitude of the observed $\lambda$EP density shifts are consistent with the geometric argument presented above.

\bibliographystyle{aipnum4-1.bst}
\bibliography{refs}

\end{document}